\documentclass[11pt]{article}
\usepackage{amssymb,amsmath,amsfonts}
\usepackage{graphicx}
\usepackage{graphics}
\usepackage{eepic,epsfig}

1.60cm \makeatletter \@addtoreset{equation}{section}

\makeatother

\begin{document}

\title{Finite temperature fermionic charge and current densities\\
induced by a cosmic string with magnetic flux}
\author{A. Mohammadi $^{1}$\thanks{%
E-mail: a.mohammadi@fisica.ufpb.br},\, E. R. Bezerra de Mello$^{1}$\thanks{%
E-mail: emello@fisica.ufpb.br} ,\, A. A. Saharian$^{2}$\thanks{%
E-mail: saharian@ysu.am} \\
\\
\textit{$^{1}$Departamento de F\'{\i}sica, Universidade Federal da Para\'{\i}%
ba}\\
\textit{58.059-970, Caixa Postal 5.008, Jo\~{a}o Pessoa, PB, Brazil}\vspace{%
0.3cm}\\
\textit{$^2$Department of Physics, Yerevan State University,}\\
\textit{1 Alex Manoogian Street, 0025 Yerevan, Armenia}}
\maketitle

\begin{abstract}
We investigate the finite temperature expectation values of the charge and
current densities for a massive fermionic field with nonzero chemical
potential, $\mu $, in the geometry of a straight cosmic string with a
magnetic flux running along its axis. These densities are decomposed into
the vacuum expectation values and contributions coming from the particles
and antiparticles. The charge density is an even periodic function of the
magnetic flux with the period equal to the quantum flux and an odd function
of the chemical potential. The only nonzero component of the current density
corresponds to the azimuthal current. The latter is an odd periodic function
of the magnetic flux and an even function of the chemical potential. At high
temperatures, the parts in the charge density and azimuthal current induced
by the planar angle deficit and magnetic flux are exponentially small. The
asymptotic behavior at low temperatures crucially depends whether the value $%
|\mu |$ is larger or smaller than the mass of the field quanta, $m$. For $%
|\mu |<m$ the charge density and the contributions into the azimuthal
current coming from the particles and antiparticles are exponentially
suppressed at low temperatures. In the case $|\mu |>m$, the charge and
current densities receive two contributions coming from the vacuum
expectation values and from particles or antiparticles (depending on the
sign of chemical potential). At large distances from the string the latter
exhibits a damping oscillatory behavior with the amplitude inversely
proportional to the square of the distance.
\end{abstract}

\bigskip


\bigskip

\section{Introduction}

In 1973, Nielsen and Olesen proposed a theoretical model comprised by Higgs
and gauge fields, that produces linear topological defect carrying a
magnetic flux named `vortex' \cite{Nielsen}. A few years later, Garfinkle
investigated the influence of the vortex on the geometry of the spacetime
\cite{Garfinkle}. Coupling the energy-momentum tensor associated with the
system to the Einstein equations, he found static cylindrically symmetric
solutions. The author also showed that, asymptotically, the spacetime around
the vortex is a Minkowski one minus a wedge. The core of the vortex has a
nonzero thickness and a magnetic flux through it. Two years later, Linet
\cite{Linet1} obtained exact solutions for the complete set of differential
equations, for some specific conditions. The author showed that the
structure of the respective spacetime corresponds to a conical one, with the
conicity parameter being expressed in terms of the energy per unit length of
the vortex.

The cosmic strings are among the most important class of linear topological
defects with the conical geometry outside the core \cite{Vile94}. The
formation of this type of topologically stable structures during the
cosmological expansion is predicted in most interesting models of high
energy physics. They have a number of interesting observable consequences,
the detection of which would provide an important link between cosmology and
particle physics. In the eighties of last century, the interest to cosmic
strings was mainly due to the fact that they provide a mechanism for the
large-scale structure formation in the universe which is alternative to the
one based on inflationary paradigm. Though the recent observations of the
cosmic microwave background radiation have ruled out the scenario in which
the cosmic strings seed the primordial density perturbations, they are
sources of a number of interesting effects which include the generation of
gravitational waves, high energy cosmic rays, and gamma ray bursts. Among
the other signatures of the cosmic strings, we mention here the
gravitational lensing, the creation of small non-Gaussianities in the cosmic
microwave background and the influence on the corresponding tensor-modes.
The recent progress in models with large compact dimensions and large warp
factors have shown that the fundamental strings can act as cosmic strings.
In particular, a new formation mechanism for cosmic strings is proposed in
the framework of brane inflation \cite{Sara02}. The cosmic string type
conical defects appear also in condensed matter systems such as crystals,
liquid crystals and quantum liquids (see, for example, Ref. \cite{Nels02}).

In quantum field theory, the nontrivial spatial topology due to the presence
of a cosmic string causes a number of interesting physical effects. In
particular, many authors have considered the vacuum polarization for scalar,
fermionic and vector fields induced by a planar angle deficit. Among the
main local characteristics of the vacuum state, the expectation values of
the field squared and of the energy-momentum tensor have been investigated.
In addition to the deficit angle parameter, the physical origin of a cosmic
string is characterized by the gauge field flux parameter describing a
magnetic flux running along the string core. The latter induces additional
polarization effects for charged fields \cite{Dowk87}-\cite{Site12}. Though
the gauge field strength vanishes outside the string core, the nonvanishing
vector potential leads to Aharonov-Bohm-like effects on scattering cross
sections and on particle production rates around the cosmic string \cite%
{Alfo89}.

For charged fields, the magnetic flux along the string core induces nonzero
vacuum expectation value of the current density. The latter, in addition to
the expectation values of the field squared and the energy-momentum tensor,
is among the most important local characteristics of the vacuum state for
quantum fields. The expectation value of the current density acts as a
source in semiclassical Maxwell equations and plays a crucial role in
modeling a self-consistent dynamics involving the electromagnetic field. The
azimuthal current density for scalar and fermionic fields, induced by a
magnetic flux in the geometry of a straight cosmic string, has been
investgated in \cite{Srir01}-\cite{Brag14}. The fermionic current induced by
a magnetic flux in a $(2+1)$-dimensional conical spacetime with a circular
boundary has been analyzed in \cite{Beze10b}. The compactification of the
cosmic string along its axis may lead to the appearance of the axial current
density \cite{Beze13,Brag14} (for the vacuum expectation value of the
current density in models with compact dimensions see \cite{Bell10} in the
case of flat background geometry and \cite{Bell13,Beze14} for de Sitter and
anti-de Sitter bulks). The generalization of the corresponding results for
the fermionic case to a cosmic string in the background of de Sitter
spacetime is given in \cite{Moha14}. All these considerations of the current
density around a cosmic string deal with the idealized geometry where the
string is assumed to have zero thickness. Realistic cosmic strings have
internal structure, characterized by the core radius determined by the
symmetry breaking scale at which it is formed. The vacuum expectation value
of the current density for a massive charged scalar field in the geometry of
a cosmic string with a general cylindrically symmetric core of a finite
support is investigated in \cite{Beze14b}. In the corresponding model, the
core encloses a gauge field flux directed along the string axis with an
arbitrary radial distribution.

Continuing in this line of investigations, in the present paper we consider
the effects of the finite temperature and nonzero chemical potential on the
expectation values of the charge and current densities for a massive
fermionic field in the geometry of a straight cosmic string for arbitrary
values of the planar angle deficit (for combined effects of the finite
temperature and nontrivial spatial topology on the expectation values of the
charge and current densities for scalar and fermionic fields in models with
toroidally compactified dimensions see \cite{Beze13T,Bell14T}). This is an
important topic, since for a cosmic string in the early stages of the
cosmological expansion the typical state of a quantum field is a state
containing particles in thermal equilibrium. The finite temperature
expectation value of the energy density for a massless scalar field around a
cosmic string in the absence of the magnetic flux is derived in \cite{Davi88}
for integer values of the parameter $q=2\pi /\phi _{0}$, where $2\pi -\phi
_{0}$ is the planar angle deficit. The expectation value of a renormalized
energy-momentum tensor for a general case of the parameter $q\geqslant 1$
has been considered in \cite{Line92} for a conformally coupled massless
scalar field and in \cite{Frol95} for a general case of a curvature coupling
parameter. Guimar\~{a}es \cite{Guim95} has extended the corresponding
results for a magnetic flux cosmic string assuming that $q<2$. The thermal
average of the energy-momentum tensor for massless fermions has been
investigated in \cite{Line96}, again, under the assumption $q<2$.

This paper is structured as follows. In the next section we describe the
background geometry associated with the spacetime under consideration and
construct a complete set of normalized positive- and negative-energy
fermionic mode functions. By using these functions, the thermal average of
the charge density is investigated in section \ref{sec:Charge}. Various
asymptotic limits are considered in detail, including the low- and
high-temperature asymptotics. The only nonzero component of the thermal
average for the current density corresponds to the azimuthal current. The
corresponding expression is derived in section \ref{sec:Current}. The
expectation value is decomposed into the vacuum expectation value and the
contributions coming from particles and antiparticles. The behavior of the
current density in the asymptotic regions of the parameters is discussed.
The main results of the paper are summarized in section \ref{sec:Conc}.

\section{Geometry and the fermionic modes}

\label{sec1}

The background geometry corresponding to a straight cosmic string lying
along the $z$-axis can be written through the line element
\begin{equation}
ds^{2}=dt^{2}-dr^{2}-r^{2}d\phi ^{2}-dz{}^{2}\ ,  \label{ds21}
\end{equation}%
where $r\geqslant 0$, $0\leqslant \phi \leqslant \phi _{0}=2\pi /q$, $%
-\infty <t<+\infty $. The parameter $q\geqslant 1$ codifies the planar angle
deficit and is related to the linear mass density of the string $\mu _{0}$
as $q^{-1}=1-4\mu _{0}$. In the presence of an external electromagnetic
field with the vector potential $A_{\mu }$, the dynamics of a massive
charged spinor field in curved spacetime is described by the Dirac equation,
\begin{equation}
(i\gamma ^{\mu }{\mathcal{D}}_{\mu }-m)\psi =0\ ,\ {\mathcal{D}}_{\mu
}=\partial _{\mu }+\Gamma _{\mu }+ieA_{\mu },  \label{Direq}
\end{equation}%
where $\gamma ^{\mu }$ are the Dirac matrices in curved spacetime and $%
\Gamma _{\mu }$ are the spin connections. For the geometry at hand the gamma
matrices can be taken in the form (see, for instance, \cite{Beze13})
\begin{equation}
\gamma ^{0}=\left(
\begin{array}{cc}
1 & 0 \\
0 & -1%
\end{array}%
\right) ,\;\gamma ^{l}=\left(
\begin{array}{cc}
0 & \rho ^{l} \\
-\rho ^{l} & 0%
\end{array}%
\right) ,  \label{gamcurved}
\end{equation}%
where the $2\times 2$ matrices $\rho ^{l}$ are
\begin{equation}
\rho ^{1}=\left(
\begin{array}{cc}
0 & e^{-iq\phi } \\
e^{iq\phi } & 0%
\end{array}%
\right) \ ,\ \rho ^{2}=-\frac{i}{r}\left(
\begin{array}{cc}
0 & e^{-iq\phi } \\
-e^{iq\phi } & 0%
\end{array}%
\right) \ ,\ \rho ^{3}=\left(
\begin{array}{cc}
1 & 0 \\
0 & -1%
\end{array}%
\right) \ .  \label{betl}
\end{equation}%
The only nonzero component of the spin connection corresponds to $\mu =2$:%
\begin{equation}
\Gamma _{\mu }=i\frac{q-1}{2}\left(
\begin{array}{cc}
\rho ^{3} & 0 \\
0 & \rho ^{3}%
\end{array}%
\right) \delta _{\mu }^{2},  \label{Gam}
\end{equation}%
and in the Dirac equation one has $\gamma ^{\mu }\Gamma _{\mu }=(1-q)\gamma
^{1}/(2r)$.

We shall admit the existence of a gauge field with the constant vector
potential as
\begin{equation}
A_{\mu }=(0,0,A_{2},0)\ .  \label{Amu}
\end{equation}%
The azimuthal component $A_{2}$ is related to an infinitesimal thin magnetic
flux, $\Phi $, running along the string, as $A_{2}=-q\Phi /(2\pi )$.
Although the magnetic field strength corresponding to the vector potential (%
\ref{Amu}) vanishes, the magnetic flux along the cosmic string leads to
Aharonov-Bohm-like effects on the expectation values of physical
observables. In particular, as it will be shown below, it induces a nonzero
expectation value of the current density.

Here, we are interested in the effects of the presence of the cosmic string
and magnetic flux on the expectation values of the charge and current
densities assuming that the field is in thermal equilibrium at finite
temperature $T$. We shall use the same procedure as in \cite{Bell14T} to
evaluate these physical observables at finite temperature. The standard form
of the density matrix for the thermodynamical equilibrium distribution at
temperature $T$ is
\begin{equation}
\hat{\rho}=Z^{-1}e^{-\beta (\hat{H}-\mu ^{\prime }\hat{Q})},\;\beta =1/T,
\label{rho}
\end{equation}%
where $\hat{H}$ is the Hamilton operator, $\hat{Q}$ denotes a conserved
charge and $\mu ^{\prime }$ is the corresponding chemical potential. The
grand canonical partition function $Z$ is given by
\begin{equation}
Z=\mathrm{tr}[e^{-\beta (\hat{H}-\mu ^{\prime }\hat{Q})}].  \label{PartFunc}
\end{equation}

Let $\{\psi _{\sigma }^{(+)},\psi _{\sigma }^{(-)}\}$ be a complete set of
normalized positive- and negative-energy solutions of (\ref{Direq}),
specified by a set of quantum numbers $\sigma $. In order to evaluate the
fermionic current densities we expand the field operator as
\begin{equation}
\psi =\sum_{\sigma }[\hat{a}_{\sigma }\psi _{\sigma }^{(+)}+\hat{b}_{\sigma
}^{+}\psi _{\sigma }^{(-)}]\ ,  \label{psiexp}
\end{equation}%
where $\hat{a}_{\sigma }$ and $\hat{b}_{\sigma }^{+}$ represent the
annihilation and creation operators corresponding to particles and
antiparticles respectively, and use the relations%
\begin{eqnarray}
\mathrm{tr}[\hat{\rho}\hat{a}_{\sigma }^{+}\hat{a}_{\sigma ^{\prime }}] &=&%
\frac{\delta _{\sigma \sigma ^{\prime }}}{e^{\beta (\varepsilon _{\sigma
}^{(+)}-\mu )}+1},  \notag \\
\mathrm{tr}[\hat{\rho}\hat{b}_{\sigma }^{+}\hat{b}_{\sigma ^{\prime }}] &=&%
\frac{\delta _{\sigma \sigma ^{\prime }}}{e^{\beta (\varepsilon _{\sigma
}^{(-)}+\mu )}+1},  \label{traa}
\end{eqnarray}%
where $\mu =e\mu ^{\prime }$ and $\pm \varepsilon _{\sigma }^{(\pm )}$ with $%
\varepsilon _{\sigma }^{(\pm )}>0$, are the energies corresponding to the
modes $\psi _{\sigma }^{(\pm )}$.

The expectation value of the fermionic current density is given by
\begin{equation}
\left\langle j^{\nu }\right\rangle =e\,\mathrm{tr}[\hat{\rho}\bar{\psi}%
(x)\gamma ^{\nu }\psi (x)].  \label{C}
\end{equation}%
Substituting the expansion (\ref{psiexp}) and using the relations (\ref{traa}%
), the current density is decomposed as
\begin{equation}
\left\langle j^{\nu }\right\rangle =\left\langle j^{\nu }\right\rangle
_{0}+\sum_{\chi =+,-}\left\langle j^{\nu }\right\rangle _{\chi },  \label{C1}
\end{equation}%
where
\begin{equation}
\left\langle j^{\nu }\right\rangle _{0}=e\sum_{\sigma }\bar{\psi}_{\sigma
}^{(-)}(x)\gamma ^{\nu }\psi _{\sigma }^{(-)}(x),  \label{Cvev}
\end{equation}%
is the vacuum expectation value and
\begin{equation}
\left\langle j^{\nu }\right\rangle _{\pm }=\pm e\sum_{\sigma }\frac{\bar{\psi%
}_{\sigma }^{(\pm )}\gamma ^{\nu }\psi _{\sigma }^{(\pm )}}{e^{\beta
(\varepsilon _{\sigma }^{(\pm )}\mp \mu )}+1}.  \label{jpm}
\end{equation}%
Here, $\left\langle j^{\nu }\right\rangle _{\pm }$ is the part in the
expectation value coming from the particles for the upper sign and from the
antiparticles for the lower sign.

As is seen, for the evaluation of the current density we need a complete set
of fermionic modes. In Ref. \cite{Beze13}, it has been shown that the
positive- and negative-energy fermionic mode-functions are uniquely
specified by the set of quantum numbers $\sigma =(\lambda ,k,j,s)$ with%
\begin{equation}
\lambda \geqslant 0,\;-\infty <k<+\infty ,\;j=\pm 1/2,\pm 3/2,\ldots
,\;s=\pm 1.  \label{range}
\end{equation}%
These functions are expressed as
\begin{equation}
\psi _{\sigma }^{(\pm )}(x)=C_{\sigma }^{(\pm )}e^{\mp iEt+ikz+iqj\phi
}\left(
\begin{array}{c}
J_{\beta _{j}}(\lambda r)e^{-iq\phi /2} \\
sJ_{\beta _{j}+\epsilon _{j}}(\lambda r)e^{iq\phi /2} \\
\pm \frac{k-is\epsilon _{j}\lambda }{E\pm m}J_{\beta _{j}}(\lambda
r)e^{-iq\phi /2} \\
\mp s\frac{k-is\lambda \epsilon _{j}}{E\pm m}J_{\beta _{j}+\epsilon
_{j}}(\lambda r)e^{iq\phi /2}%
\end{array}%
\right) \ ,  \label{psi+n}
\end{equation}%
where $J_{\nu }(x)$ is the Bessel function,
\begin{equation}
E=\varepsilon _{\sigma }^{(\pm )}=\sqrt{\lambda ^{2}+k^{2}+m^{2}},  \label{E}
\end{equation}%
and
\begin{equation}
\beta _{j}=q|j+\alpha |-\epsilon _{j}/2\ ,\;\alpha =eA_{2}/q=-\Phi /\Phi
_{0},  \label{betaj}
\end{equation}%
with $\epsilon _{j}=\mathrm{sgn}(j+\alpha )$ and $\Phi _{0}=2\pi /e$ being
the flux quantum. The wave functions (\ref{psi+n}) are eigenfunctions of the
projection of total angular momentum operator along the cosmic string,
\begin{equation}
\hat{J}_{3}\psi _{\sigma }^{(\pm )}=\left( -\frac{i}{q}(\partial
_{2}+ieA_{2})+\frac{1}{2}\left(
\begin{array}{cc}
\rho ^{3} & 0 \\
0 & \rho ^{3}%
\end{array}%
\right) \right) \psi _{\sigma }^{(\pm )}=(j+\alpha )\psi _{\sigma }^{(\pm )},
\label{J3}
\end{equation}%
with the eigenvalues $j+\alpha $.

The constants $C_{\sigma }^{(\pm )}$ in (\ref{psi+n}) are determined by the
orthonormalization condition
\begin{equation}
\int d^{3}x\sqrt{\gamma }\ (\psi _{\sigma }^{(\pm )})^{\dagger }\psi
_{\sigma ^{\prime }}^{(\pm )}=\delta _{\sigma \sigma ^{\prime }}\ ,
\label{normcond}
\end{equation}%
where $\gamma $ is the determinant of the spatial metric tensor. The delta
symbol on the right-hand side is understood as the Dirac delta function for
continuous quantum numbers ($\lambda ,k$) and the Kronecker delta for
discrete ones ($j,s$). From (\ref{normcond}) one obtains
\begin{equation}
|C_{\sigma }^{(\pm )}|^{2}=\frac{q\lambda (E\pm m)}{16\pi ^{2}E}\ .
\label{C+}
\end{equation}%

In defining the fermionic mode functions (\ref{psi+n}) we have
imposed the regularity condition at the cone apex. It is well known
that for an idealized zero thickness magnetic flux the theory of von
Neumann deficiency indices leads to a one-parameter family of
boundary conditions \cite{Sous89}. In addition to the regular modes,
these conditions, in general, allow normalizable irregular modes.
The expectation values of the charge and current densities for
general boundary conditions are evaluated in a way similar to that
described below. The contribution of the regular modes is the same
for all boundary conditions and the expressions differ by the
contributions coming from the irregular modes. Note that in a recent
investigation of the induced fermionic current for a massless Dirac
field in (2+1) dimensions \cite{Jack09}, the authors impose the
regularity condition. The corresponding result agrees with that for
a finite radius solenoid, assuming that fermions cannot penetrate
the region of nonzero magnetic flux.

The fermionic charge and current densities at zero temperature, $%
\left\langle j^{\mu }\right\rangle _{0}$, have been investigated in \cite%
{Beze13} for zero chemical potential. Therefore, we will be mainly concerned
with the contributions from particles and antiparticles provided by the
second term in the right-hand side of (\ref{C1}).

\section{Charge density}

\label{sec:Charge}

We start with the charge density corresponding to the $\nu =0$ component in (%
\ref{C1}). As it has been shown in \cite{Beze13}, the vacuum expectation
value of the charge density vanishes, $\left\langle j^{0}\right\rangle
_{0}=0 $. Substituting the mode functions (\ref{psi+n}) into (\ref{jpm}),
for the contributions coming from the particles and antiparticles we get
\begin{equation}
\left\langle j^{0}\right\rangle _{\pm }=\pm \frac{eq}{8\pi ^{2}}\sum_{\sigma
}\lambda \frac{J_{\beta _{j}}^{2}(\lambda r)+J_{\beta _{j}+\epsilon
_{j}}^{2}(\lambda r)}{e^{\beta (E\mp \mu )}+1},  \label{j0pm}
\end{equation}%
where we use the notation
\begin{equation}
\sum_{\sigma }=\int_{-\infty }^{+\infty }dk\int_{0}^{\infty }d\lambda \
\sum_{s=\pm 1}\sum_{j}\ .  \label{Sumsig}
\end{equation}%
Here and in what follows%
\begin{equation}
\sum_{j}=\sum_{j=\pm 1/2,\pm 3/2,\cdots }.  \label{Sumj}
\end{equation}%
As is seen, in the case $\mu =0$ the contributions from the particles and
antiparticles cancel each other and the total charge density,%
\begin{equation}
\left\langle j^{0}\right\rangle =\left\langle j^{0}\right\rangle
_{+}+\left\langle j^{0}\right\rangle _{-},  \label{j0tot}
\end{equation}%
is zero. For $\mu >0$ ($\mu <0$) the particles (antiparticles) dominate and $%
\left\langle j^{0}\right\rangle /e>0$ ($\left\langle j^{0}\right\rangle /e<0$%
). From (\ref{j0pm}) it follows that the charge density is an even periodic
function of the parameter $\alpha $ with the period equal to 1. This means
that the charge density is a periodic function of the magnetic flux with the
period equal to the quantum flux. If we present this parameter as%
\begin{equation}
\alpha =n_{0}+\alpha _{0},\;|\alpha _{0}|\leqslant 1/2,  \label{alf0}
\end{equation}%
with $n_{0}$ being an integer, then the current density depends on $\alpha
_{0}$ alone.

For the further transformation of the expression (\ref{j0pm}), first we
consider the case $|\mu |<$ $m$. With this assumption, by using the
expansion
\begin{equation}
(e^{y}+1)^{-1}=-\sum_{n=1}^{\infty }(-1)^{n}e^{-ny}\ ,  \label{Expans}
\end{equation}%
after the summation over $s$, we obtain
\begin{eqnarray}
\left\langle j^{0}\right\rangle _{\pm } &=&\mp \frac{eq}{2\pi ^{2}}%
\sum_{n=1}^{\infty }(-1)^{n}e^{\pm n\beta \mu }\sum_{j}\int_{0}^{\infty }dk
\notag \\
&&\times \int_{0}^{\infty }d\lambda \,\lambda \lbrack J_{\beta
_{j}}^{2}(\lambda r)+J_{\beta _{j}+\epsilon _{j}}^{2}(\lambda r)]e^{-n\beta
E}\ .  \label{j0pm1}
\end{eqnarray}%
As the next step, we use the integral representation%
\begin{equation}
e^{-n\beta E}=\frac{n\beta }{\sqrt{\pi }}\int_{0}^{\infty
}ds\,s^{-2}e^{-(\lambda ^{2}+k^{2}+m^{2})s^{2}-n^{2}\beta ^{2}/4s^{2}},
\label{IntRep2}
\end{equation}%
and the relation \cite{Grad}
\begin{equation}
\int_{0}^{\infty }d\lambda \ \lambda e^{-s^{2}\lambda ^{2}}[J_{\beta
_{j}}^{2}(\lambda r)+J_{\beta _{j}+\epsilon _{j}}^{2}(\lambda r)]=\frac{%
e^{-x}}{2s^{2}}\left[ I_{\beta _{j}}(x)+I_{\beta _{j}+\epsilon _{j}}(x)%
\right] \ ,  \label{Int-reg}
\end{equation}%
with $x=r^{2}/(2s^{2})$ and $I_{\nu }(x)$ being the modified Bessel
function. After integration over $k$, we get
\begin{eqnarray}
\left\langle j^{0}\right\rangle _{\pm } &=&\mp \frac{eq\beta }{4\pi ^{2}r^{4}%
}\ \sum_{n=1}^{\infty }(-1)^{n}ne^{\pm n\beta \mu }\int_{0}^{\infty }dx\,x
\notag \\
&&\times F(q,\alpha _{0},x)e^{-m^{2}r^{2}/2x-(1+n^{2}\beta ^{2}/2r^{2})x}\ ,
\label{j0pm2}
\end{eqnarray}%
where the notation%
\begin{equation}
F(q,\alpha _{0},x)=\sum_{j}\ \left[ I_{\beta _{j}}(x)+I_{\beta _{j}+\epsilon
_{j}}(x)\right] ,  \label{Fq}
\end{equation}%
is introduced

An alternative representation for the charge density is obtained by making
use of the integral representation for the series $\sum_{j}I_{\beta _{j}}(x)$
derived in \cite{Beze10b}. With this representation, the function (\ref{Fq})
is expressed as%
\begin{eqnarray}
F(q,\alpha _{0},x) &=&\frac{4}{q}\left[ \frac{e^{x}}{2}%
+\sum_{k=1}^{[q/2]}(-1)^{k}c_{k}\cos \left( 2\pi k\alpha _{0}\right)
e^{x\cos (2\pi k/q)}\right.  \notag \\
&&\left. +\frac{q}{\pi }\int_{0}^{\infty }dy\frac{h(q,\alpha _{0},2y)\sinh y%
}{\cosh (2qy)-\cos (q\pi )}e^{-x\cosh {(2y)}}\right] \ ,  \label{Fq1}
\end{eqnarray}%
where $[q/2]$ means the integer part of $q/2$ and the notation%
\begin{equation}
h(q,\alpha _{0},x)=\sum_{\chi =\pm 1}\cos \left[ \left( 1/2+\chi \alpha
_{0}\right) q\pi \right] \sinh \left[ \left( 1/2-\chi \alpha _{0}\right) qx%
\right] ,  \label{h}
\end{equation}%
is assumed. Here and in what follows we use the notations%
\begin{equation}
c_{k}=\cos {(\pi k/q),\;s}_{k}=\sin {(\pi k/q).}  \label{cksk}
\end{equation}%
Substituting (\ref{Fq1}) into (\ref{j0pm2}), after integration over $x$, we
find the expression%
\begin{eqnarray}
\left\langle j^{0}\right\rangle _{\pm } &=&\left\langle j^{0}\right\rangle _{%
\mathrm{M}\pm }\mp \frac{2em^{4}\beta }{\pi ^{2}}\ \sum_{n=1}^{\infty
}(-1)^{n}ne^{\pm n\beta \mu }  \notag \\
&&\times \left[ \sum_{k=1}^{[q/2]}(-1)^{k}c_{k}\cos \left( 2\pi k\alpha
_{0}\right) f_{2}(m\beta s_{n}(r/\beta ,k/q))\right.  \notag \\
&&\left. +\frac{q}{\pi }\int_{0}^{\infty }dy\frac{\sinh \left( y\right)
h(q,\alpha _{0},2y)}{\cosh (2qy)-\cos (q\pi )}f_{2}(m\beta c_{n}(r/\beta ,y))%
\right] ,  \label{j0pm3}
\end{eqnarray}%
where%
\begin{equation}
\left\langle j^{0}\right\rangle _{\mathrm{M}\pm }=\mp \frac{e\beta m^{4}}{%
\pi ^{2}}\sum_{n=1}^{\infty }(-1)^{n}ne^{\pm n\beta \mu }f_{2}(nm\beta ),
\label{j0pmM}
\end{equation}%
is the corresponding charge density in Minkowski spacetime in the absence of
the magnetic flux and the cosmic string ($\alpha _{0}=0$, $q=1$). Here we
have introduced the notations%
\begin{equation}
f_{\nu }(x)=x^{-\nu }K_{\nu }(x),  \label{fnu1}
\end{equation}%
with $K_{\nu }(x)$ being the MacDonald function and
\begin{eqnarray}
s_{n}(x,y) &=&\sqrt{n^{2}+4x^{2}\sin ^{2}(\pi y)},  \notag \\
c_{n}(x,y) &=&\sqrt{n^{2}+4x^{2}\cosh ^{2}y}\ .  \label{sncn}
\end{eqnarray}

For the total charge density one gets%
\begin{eqnarray}
\left\langle j^{0}\right\rangle &=&-\frac{4em^{4}\beta }{\pi ^{2}}\
\sum_{n=1}^{\infty }(-1)^{n}n\sinh (n\beta \mu )\left[ \frac{1}{2}%
f_{2}(m\beta n)\right.  \notag \\
&&+\sum_{k=1}^{[q/2]}(-1)^{k}c_{k}\cos \left( 2\pi k\alpha _{0}\right)
f_{2}(m\beta s_{n}(r/\beta ,k/q))  \notag \\
&&\left. +\frac{q}{\pi }\int_{0}^{\infty }dy\frac{h(q,\alpha _{0},2y)\sinh y%
}{\cosh (2qy)-\cos (q\pi )}f_{2}(m\beta c_{n}(r/\beta ,y))\right] ,
\label{j0}
\end{eqnarray}%
where the contribution coming from the first term in the square brackets
corresponds to the charge density in Minkowski spacetime, $\left\langle
j^{0}\right\rangle _{\mathrm{M}}=\left\langle j^{0}\right\rangle _{\mathrm{M}%
+}+\left\langle j^{0}\right\rangle _{\mathrm{M}-}$. Note that the expression
(\ref{j0}) is also valid in the case $|\mu |=m$. In the absence of the
conical defect, $q=1$, the expression (\ref{j0}) reduces to
\begin{align}
\left\langle j^{0}\right\rangle & =-\frac{2em^{4}\beta }{\pi ^{2}}%
\sum_{n=1}^{\infty }(-1)^{n}n\sinh (n\beta \mu )\left[ f_{2}(nm\beta )\right.
\notag \\
& +\left. \frac{2}{\pi }\sin \left( \alpha _{0}\pi \right) \int_{0}^{\infty
}dy\,\tanh y\sinh \left( 2\alpha _{0}y\right) f_{2}\left( m\beta
c_{n}(r/\beta ,y)\right) \right] ,  \label{j0q1}
\end{align}%
where the second term in the square brackets is induced by the magnetic flux.

Let us consider the behavior of the charge density in various asymptotic
regions of the parameters. First we consider the region near the string. For
$2|\alpha _{0}|\leqslant 1-1/q$, the charge density is finite on the string.
The corresponding limiting value is directly obtained from (\ref{j0}) by
taking $r=0$:
\begin{equation}
\left\langle j^{0}\right\rangle _{r=0}=-\frac{2em^{4}\beta }{\pi ^{2}}\left[
1+2h_{0}(q,\alpha _{0})\right] \sum_{n=1}^{\infty }(-1)^{n}n\sinh (n\beta
\mu )f_{2}(nm\beta ),  \label{j0r0n}
\end{equation}%
where we have used the notation%
\begin{equation}
h_{n}(q,\alpha _{0})=\sum_{k=1}^{[q/2]}(-1)^{k}\frac{c_{k}}{s_{k}^{n}}\cos
\left( 2\pi k\alpha _{0}\right) +\frac{q}{\pi }\int_{0}^{\infty }dy\ \frac{%
h(q,\alpha _{0},2y)\sinh \left( y\right) \cosh ^{-n}y}{\cosh (2qy)-\cos
(q\pi )}.  \label{hnq}
\end{equation}%
In the case $2|\alpha _{0}|>1-1/q$ the charge density diverges on the
string. This divergence comes from the integral term in (\ref{j0}). In order
to find the leading term in the corresponding asymptotic expansion over the
distance from the string, we note that for small $r$ the dominant
contribution to the integral comes from large values of $y$. By using the
corresponding asymptotic, the integral over $y$ is expressed in terms of the
MacDonald function and to the leading order one gets%
\begin{equation}
\left\langle j^{0}\right\rangle \approx -\frac{eqm^{4}\beta (mr/\sqrt{2}%
)^{\left( 1-2|\alpha _{0}|\right) q-1}}{\pi ^{2}\Gamma (1/2+\left(
1/2-|\alpha _{0}|\right) q)}\sum_{n=1}^{\infty }(-1)^{n}n\sinh (n\beta \mu
)f_{3/2+\left( 1/2-|\alpha _{0}|\right) q}(nm\beta ).  \label{j0r0b}
\end{equation}%
As is seen, the divergence is integrable and, hence, the part in the total
charge per unit length of the string, induced by the cosmic string and
magnetic flux, is finite. We denote this charge by $\Delta Q$. The charge
density, induced by the cosmic string and magnetic flux, is given by the
second and third terms in the square brackets of (\ref{j0}). After
integration of this part over the radial and angular coordinates, one gets%
\begin{equation}
\Delta Q=-\frac{2em^{2}\beta }{\pi q}h_{2}(q,\alpha _{0})\
\sum_{n=1}^{\infty }(-1)^{n}n\sinh (n\beta \mu )f_{1}(m\beta n).
\label{DelQ}
\end{equation}%
For sufficiently large $r_{0}$, the total charge (per unit length of the
string) in the cylindrical volume of the radius $r_{0}$ can be written as%
\begin{equation}
Q=\frac{1}{2}\phi _{0}r_{0}^{2}\left\langle j^{0}\right\rangle _{\mathrm{M}%
}+\Delta Q.  \label{Q}
\end{equation}

Now let us consider the asymptotic behavior of the charge density at low and
high temperatures. At low temperatures, $T\ll m,r^{-1}$, and for $|\alpha
_{0}|<1/2-1/(2q)$, in the arguments of the functions $f_{2}(m\beta
s_{n}(r/\beta ,k/q))$ and $f_{2}(m\beta c_{n}(r/\beta ,y))$ in (\ref{j0}) we
can directly put $r/\beta =0$. The dominant contribution comes from the term
$n=1$ and, using the asymptotic expression of the MacDonald function for
large arguments, we find
\begin{equation}
\left\langle j^{0}\right\rangle \approx \frac{2em^{3}\mathrm{sgn}(\mu )}{%
(2\pi m\beta )^{3/2}}\frac{1+2h_{0}(q,\alpha _{0})}{e^{\beta (m-|\mu |)}}.
\label{j0T0}
\end{equation}%
For $2|\alpha _{0}|>1-1/q$, the dominant contribution to the $y$-integral in
(\ref{j0}) comes from large values of $y$. By using the corresponding
asymptotic expression for the integrand, to the leading order, one gets
\begin{equation}
\left\langle j^{0}\right\rangle \approx \frac{eqm^{3}}{(2\pi )^{3/2}}\frac{%
\mathrm{sgn}(\mu )e^{-\beta (m-|\mu |)}}{(m\beta )^{1+\left( 1/2-|\alpha
_{0}|\right) q}}\frac{(mr/\sqrt{2})^{\left( 1-2|\alpha _{0}|\right) q-1}}{%
\Gamma (1/2+\left( 1/2-|\alpha _{0}|\right) q)}.  \label{j0T0b}
\end{equation}%
Similarly to the previous case, the dominant contribution comes from the $%
n=1 $ term.

At high temperatures, the main contribution to the charge density in (\ref%
{j0}) comes from large $n$ and this representation is not convenient for the
investigation of the corresponding asymptotic. In order to find a
representation more convenient in the high temperature limit, first we make
the replacement%
\begin{equation}
(-1)^{n}n\sinh (n\beta \mu )=\frac{1}{\beta }\partial _{\mu }\cos [n(\pi
+i\beta \mu )],  \label{Repl1}
\end{equation}%
in the part induced by the cosmic string and magnetic flux and then use the
relation \cite{Bell09}
\begin{eqnarray}
&&\sum_{n=-\infty }^{+\infty }\cos (nb)f_{\nu }(m\sqrt{\beta ^{2}n^{2}+a^{2}}%
)=\frac{(2\pi )^{1/2}}{\beta m^{2\nu }}  \notag \\
&&\quad \times \sum_{n=-\infty }^{+\infty }\left[ (2\pi n+b)^{2}/\beta
^{2}+m^{2}\right] ^{\nu -1/2}f_{\nu -1/2}(a\sqrt{(2\pi n+b)^{2}/\beta
^{2}+m^{2}}),  \label{SumHighT}
\end{eqnarray}%
with $b=\pi +i\beta \mu $. In this way, we can see that%
\begin{equation}
\sum_{n=1}^{\infty }(-1)^{n}n\sinh (n\beta \mu )f_{2}(m\sqrt{\beta
^{2}n^{2}+a^{2}})=-\frac{i\pi T^{2}}{2m^{4}a}\sum_{n=-\infty }^{\infty }%
\left[ \pi \left( 2n+1\right) T+i\mu \right] e^{-ab_{n}(T)}.
\label{SumHighT2}
\end{equation}%
where%
\begin{equation}
b_{n}(T)=\sqrt{[\pi (2n+1)T+i\mu ]^{2}+m^{2}}.  \label{bn}
\end{equation}%
This gives the following representation for the charge density%
\begin{align}
\left\langle j^{0}\right\rangle & =\left\langle j^{0}\right\rangle _{\mathrm{%
M}}+\frac{ieT}{\pi r}\sum_{n=-\infty }^{+\infty }\left[ \pi (2n+1)T+i\mu %
\right]  \notag \\
& \times \left[ \sum_{k=1}^{[q/2]}(-1)^{k}\frac{c_{k}}{s_{k}}\cos \left(
2\pi k\alpha _{0}\right) e^{-2rs_{k}b_{n}(T)}\right.  \notag \\
& +\left. \frac{q}{\pi }\int_{0}^{\infty }dy\ \frac{\sinh \left( y\right)
h(q,\alpha _{0},2y)}{\cosh (2qy)-\cos (q\pi )}\frac{e^{-2rb_{n}(T)\cosh y}}{%
\cosh y}\right] .  \label{j0HighT}
\end{align}%
In deriving (\ref{j0HighT}) we have used the relation $\left[ x^{2\nu
}f_{\nu }(x)\right] ^{\prime }=-x^{2\nu -1}f_{\nu -1}(x)$ and the expression
$f_{1/2}(x)=\sqrt{\pi /2}e^{-x}/x$. Note that the series in the right-hand
side of (\ref{j0HighT}) can be written as $\sum_{n=-\infty }^{+\infty
}\cdots =2i\sum_{n=0}^{\infty }\mathrm{Im}\left[ \cdots \right] $, which
explicitly shows that the corresponding expression is real.

At high temperatures, $T\gg m,r^{-1}$, the dominant contribution in the
second term in the right-hand side of (\ref{j0HighT}) comes from the terms
with $n=0$ and $n=-1$. For $q>2$, under the condition $\sqrt{rT}{(q-{2})\gg 1%
}$ ($q$ is not too close to 2) the leading term corresponds to the $k=1$
term and we get%
\begin{equation}
\left\langle j^{0}\right\rangle \approx \left\langle j^{0}\right\rangle _{%
\mathrm{M}}-\frac{2eT^{2}}{r}\frac{{\sin [2r\mu \sin (\pi /q)]}}{e^{2\pi
rT\sin (\pi /q)}}\cot {({\pi }/q)}\cos \left( 2\pi \alpha _{0}\right) .
\label{j0HighT1}
\end{equation}%
The high-temperature asymptotic of the Minkowskian part is given by%
\begin{equation}
\left\langle j^{0}\right\rangle _{\mathrm{M}}\approx \frac{1}{3}e\mu T^{2}.
\label{j0MhighT}
\end{equation}%
For $q\leqslant 2$ the sum over $k$ in (\ref{j0HighT}) is absent. The
dominant contribution to the integral over $y$ comes from the region near
the lower limit of the integration. Assuming that $\sin (\pi q/2)\gg q/\sqrt{%
\pi rT}$ ($q$ is not too close to 2), to the leading order one finds%
\begin{equation}
\left\langle j^{0}\right\rangle \approx \left\langle j^{0}\right\rangle _{%
\mathrm{M}}+\frac{eq^{2}\sqrt{rT}}{4\pi ^{2}r^{3}e^{2\pi rT}}\frac{\sin
(2r\mu )}{\sin ^{2}(\pi q/2)}\sum_{\chi =\pm 1}\left( 1-2\chi \alpha
_{0}\right) \cos \left[ \left( 1/2+\chi \alpha _{0}\right) q\pi \right] .
\label{j0HighT2}
\end{equation}%
For the case $\sin (\pi q/2)\ll q/\sqrt{\pi rT}$, the asymptotic expression
takes the form%
\begin{equation}
\left\langle j^{0}\right\rangle \approx \left\langle j^{0}\right\rangle _{%
\mathrm{M}}-\frac{e(rT)^{3/2}}{\pi r^{3}e^{2\pi rT}}\sin (2r\mu )\cos \left(
2\pi \alpha _{0}\right) .  \label{j0HighT3}
\end{equation}%
Hence, in all cases, at high temperatures the charge density is dominated by
the Minkowskian part $\left\langle j^{0}\right\rangle _{\mathrm{M}}$ and the
contributions induced by the planar angle deficit and the magnetic flux are
exponentially small for points not too close to the string.

The representation (\ref{j0HighT}) is also convenient for the investigation
of the charge density at large distances from the string, $r\gg
T^{-1},m^{-1} $. The corresponding procedure is similar to that for the high
temperature asymptotic. The dominant contribution comes from the terms with $%
n=0$ and $n=-1$ and for $q>2$ one gets%
\begin{equation}
\left\langle j^{0}\right\rangle \approx \left\langle j^{0}\right\rangle _{%
\mathrm{M}}+\frac{2eT\cos \left( 2\pi \alpha _{0}\right) }{\pi r\tan (\pi /q)%
}\,\mathrm{Im}[\left( \pi T+i\mu \right) e^{-2rb_{0}(T)\sin (\pi /q)}],
\label{j0Larger}
\end{equation}%
where $b_{0}(T)=\sqrt{(\pi T+i\mu )^{2}+m^{2}}$. In the case $q<2$ the
corresponding asymptotic expression has the form%
\begin{equation}
\left\langle j^{0}\right\rangle \approx \left\langle j^{0}\right\rangle _{%
\mathrm{M}}-\frac{eq^{2}T}{4r\sin ^{2}(\pi q/2)}\mathrm{Im}[\frac{\left( \pi
T+i\mu \right) e^{-2rb_{0}(T)}}{\left( \pi rb_{0}(T)\right) ^{3/2}}%
]\sum_{\chi =\pm 1}\left( 1-2\chi \alpha _{0}\right) \cos \left[ \left(
1/2+\chi \alpha _{0}\right) q\pi \right] .  \label{j0Largerb}
\end{equation}%
In both cases, the parts in the charge density induced by the string and
magnetic flux are exponentially small and to the leading order one has $%
\left\langle j^{0}\right\rangle \approx \left\langle j^{0}\right\rangle _{%
\mathrm{M}}$.

\begin{figure}[tbph]
\begin{center}
\begin{tabular}{cc}
\epsfig{figure=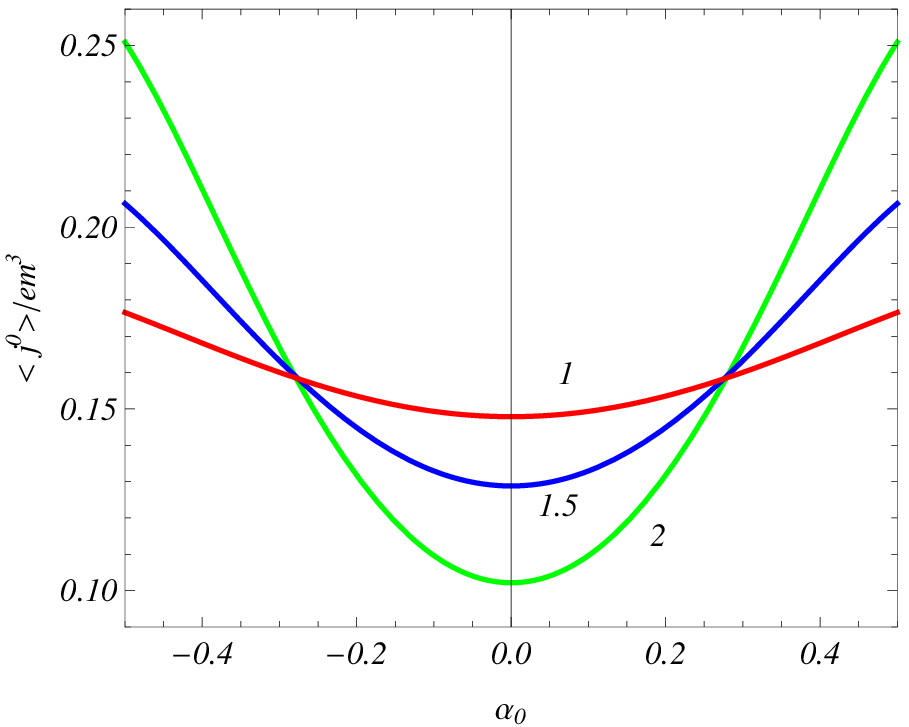,width=7.cm,height=5.5cm} & \quad %
\epsfig{figure=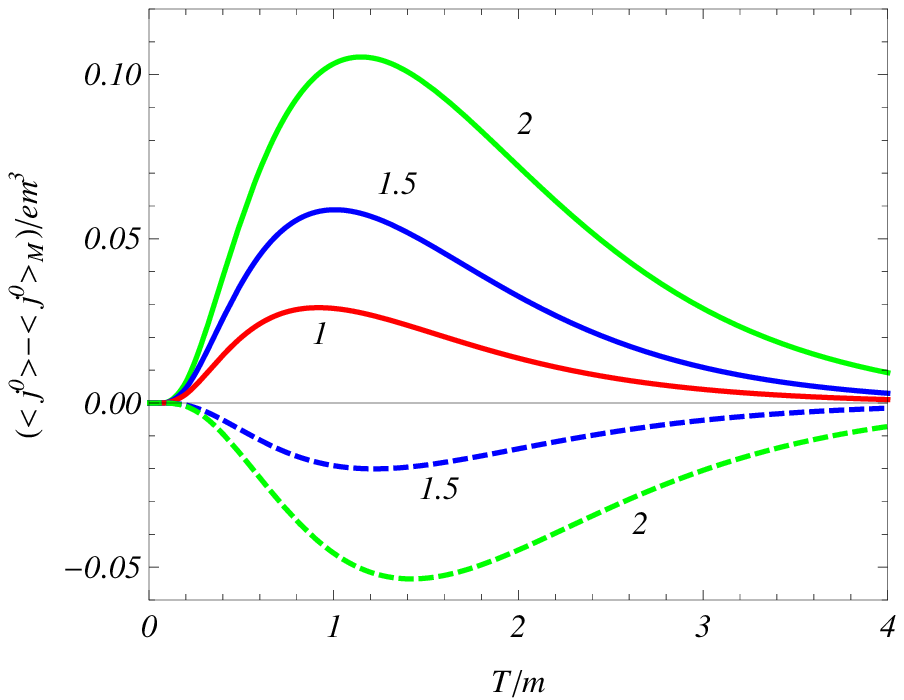,width=7.cm,height=5.5cm}%
\end{tabular}%
\end{center}
\caption{The total charge density as a function of the parameter $\protect%
\alpha _{0}$ (left panel) and the charge density induced by the string and
magnetic flux as a function of the temperature (right panel). The numbers
near the curves correspond to the values of the parameter $q$. The graphs on
the left panel are plotted for $T/m=1$, $mr=0.25 $, and $\protect\mu /m=0.5$%
. On the right panel, the full and dashed curves correspond to the values $%
\protect\alpha _{0}=1/2$ and $\protect\alpha _{0}=0$, respectively. For the
graphs on the right panel we have taken $\protect\mu /m=0.5$ and $mr=1/8$.}
\label{fig1}
\end{figure}

In figure \ref{fig1}, we have plotted the total charge density as a function
of the parameter $\alpha _{0}$ (left panel) and the charge density induced
by the string and magnetic flux as a function of the temperature (right
panel). The numbers near the curves correspond to the values of the
parameter $q$. The graphs on the left panel are plotted for $T/m=1$, $%
mr=0.25 $, and $\mu /m=0.5$. On the right panel, the full and dashed curves
correspond to the values $\alpha _{0}=1/2$ and $\alpha _{0}=0$, respectively
(note that for $q=1$, $\alpha _{0}=0$ one has $\left\langle
j^{0}\right\rangle -\left\langle j^{0}\right\rangle _{\mathrm{M}}=0$). For
the graphs on the right panel we have taken $\mu /m=0.5$ and $mr=1/8$.

In the discussion above we have assumed that $|\mu |<m$. Now we turn to the
case $|\mu |>m$. The charge density is an odd function of the chemical
potential and for definiteness we assume that $\mu >m$. In this case the
contribution of the antiparticles to the charge density is evaluated in a
way similar to that described above and the corresponding expression is
given by (\ref{j0pm3}) with the lower sign. The total charge density is
expressed as%
\begin{equation}
\left\langle j^{0}\right\rangle =\left\langle j^{0}\right\rangle _{-}+\frac{%
eq}{2\pi ^{2}}\int_{0}^{\infty }dk\int_{0}^{\infty }d\lambda \
\sum_{j}\lambda \frac{J_{\beta _{j}}^{2}(\lambda r)+J_{\beta _{j}+\epsilon
_{j}}^{2}(\lambda r)}{e^{\beta (E-\mu )}+1},  \label{j0b}
\end{equation}%
where the second term in the right-hand side is the contribution from
particles. The integrals over $k$ and $\lambda $ are divided into regions
corresponding to $\sqrt{k^{2}+\lambda ^{2}}<p_{0}$ and $\sqrt{k^{2}+\lambda
^{2}}>p_{0}$ with $p_{0}=\sqrt{\mu ^{2}-m^{2}}$ being the Fermi momentum. In
the second region we can again use the expansion (\ref{Expans}). At high
temperatures, $T\gg \mu $, the contribution of the states with $E\gg \mu $
dominates and the asymptotic expressions given above are valid.

The situation is essentially changed at low temperatures. Now, in the limit $%
T\rightarrow 0$, only the contribution of the states with $E\leqslant \mu $
survives and from (\ref{j0b}) one gets
\begin{equation}
\left\langle j^{0}\right\rangle _{T=0}=\frac{eqp_{0}^{3}}{2\pi ^{2}}\
\sum_{j}\int_{0}^{1}dx\,x\sqrt{1-x^{2}}\left[ J_{\beta
_{j}}^{2}(p_{0}rx)+J_{\beta _{j}+\epsilon _{j}}^{2}(p_{0}rx)\right] .
\label{j0bT0}
\end{equation}%
The integral in this expression is expressed in terms of the hypergeometric
function $_{1}F_{2}(a;b,c;x)$. The appearance of the nonzero charge density
at zero temperature is related to the presence of particles filling the
states with the energies $m\leqslant E\leqslant \mu $. In the case $\mu <0$
we would have antiparticles. In the absence of the planar angle deficit and
magnetic flux ($q=1$, $\alpha =0$), the summation over $j$ in (\ref{j0bT0})
gives 2 and for the charge density we obtain the standard expression:%
\begin{equation}
\left\langle j^{0}\right\rangle _{\mathrm{M},T=0}=\frac{ep_{0}^{3}}{3\pi ^{2}%
}.  \label{j0MbT0}
\end{equation}

An alternative expression for the charge density at zero temperature is
obtained from (\ref{j0bT0}) by using the relation \cite{Wats66}%
\begin{equation}
J_{\beta _{j}}^{2}(u)=\frac{1}{2\pi i}\int_{c-i\infty }^{c+i\infty }\frac{dt%
}{t}e^{t/2-u^{2}/t}I_{\beta _{j}}(u^{2}/t),  \label{JRep}
\end{equation}%
where $c$ is a positive number. With this relation, the series over $j$ is
expressed in terms of the function (\ref{Fq}). By making use of the
representation (\ref{Fq1}) for $F(q,\alpha _{0},x)$, we can see that the
contribution coming from the first term in the square brackets of (\ref{Fq1}%
) gives the charge density $\left\langle j^{0}\right\rangle _{\mathrm{M}%
,T=0} $. In the remaining part, the integral over $t$ is expressed in terms
of the Bessel function with the help of the formula%
\begin{equation}
\frac{1}{2\pi i}\int_{c-i\infty }^{c+i\infty }\frac{dt}{t}%
e^{t/2-2b^{2}x^{2}/t}=J_{0}(2bx).  \label{J0Rep}
\end{equation}%
This formula is obtained by using the integral representation of the Bessel
function from \cite{Wats66} (page 176, formula (1)) deforming the
integration contour in the complex plane $t$. The remaining integral is of
the form $\int_{0}^{1}dx\,x\sqrt{1-x^{2}}J_{0}(2bx)$. The latter is
expressed in terms of the function $J_{3/2}(2b)$ and for the charge density
we find the representation%
\begin{eqnarray}
\left\langle j^{0}\right\rangle _{T=0} &=&\left\langle j^{0}\right\rangle _{%
\mathrm{M},T=0}-\frac{ep_{0}}{2\pi ^{2}r^{2}}\ \left[
\sum_{k=1}^{[q/2]}(-1)^{k}\frac{c_{k}}{s_{k}^{2}}\cos \left( 2\pi k\alpha
_{0}\right) g(2p_{0}rs_{k})\right.  \notag \\
&&\left. +\frac{q}{\pi }\int_{0}^{\infty }dy\frac{\sinh \left( y\right)
h(q,\alpha _{0},2y)}{\cosh (2qy)-\cos (q\pi )}\frac{g(2p_{0}r\cosh y)}{\cosh
^{2}y}\right] ,  \label{j0bT01}
\end{eqnarray}%
with the notation%
\begin{equation}
g(x)=\cos x-\frac{\sin x}{x}.  \label{gx}
\end{equation}%
On the string, for $2|\alpha _{0}|<1-1/q$ one has%
\begin{equation}
\left\langle j^{0}\right\rangle _{T=0,r=0}=\frac{ep_{0}^{3}}{3\pi ^{2}}\left[
1+\ 2h_{0}(q,\alpha _{0})\right] .  \label{j0br0}
\end{equation}%
In the case $2|\alpha _{0}|>1-1/q$, the asymptotic behavior near the string
is investigated in a way similar to that we have described for $|\mu |<m$.
The dominant contribution comes from the last term in the square brackets of
(\ref{j0bT01}) and we get%
\begin{equation}
\left\langle j^{0}\right\rangle _{T=0}\approx \left\langle
j^{0}\right\rangle _{\mathrm{M},T=0}+\frac{eqp_{0}^{3}}{2\pi ^{2}}\frac{%
\left[ \left( 1-2|\alpha _{0}|\right) q+2\right] ^{-1}}{\Gamma (1+\left(
1-2|\alpha _{0}|\right) q)}\left( p_{0}r\right) ^{\left( 1-2|\alpha
_{0}|\right) q-1}.  \label{j0br0b}
\end{equation}%
In this case the charge density diverges on the string and near the string
the part induced by the planar angle deficit and magnetic flux dominates the
Minkowskian part. Note that the divergence is integrable and the
corresponding total charge is finite. At large distances, the charge density
induced by the string and magnetic flux exhibits an oscillatory behavior
with the amplitude damping as $1/r^{2}$. We just want to point out that in
the case $|\mu |<m$ we had an exponential suppression.

\begin{figure}[tbph]
\begin{center}
\begin{tabular}{cc}
\epsfig{figure=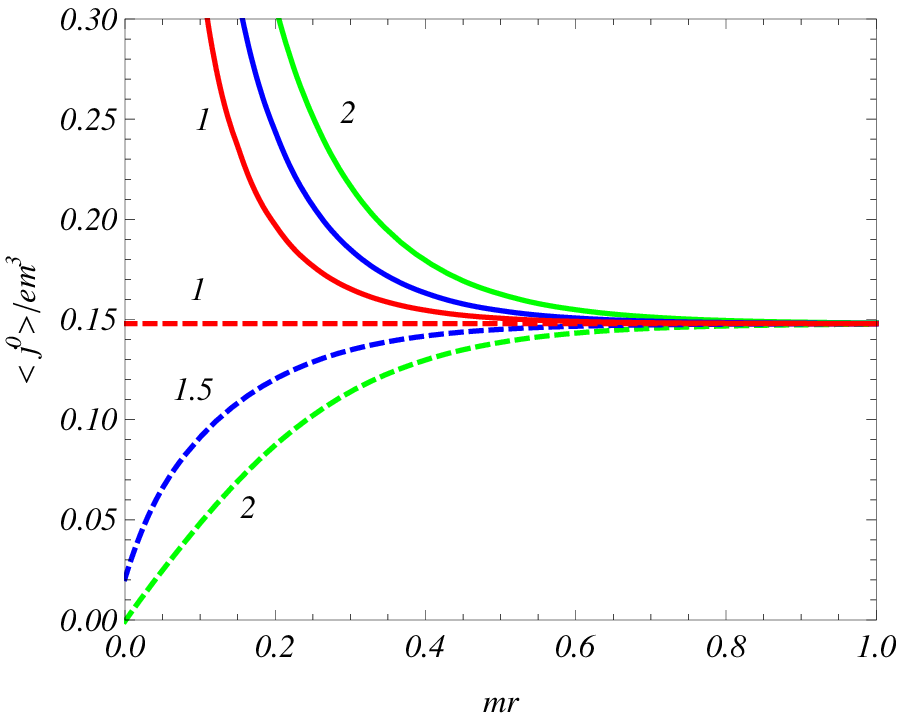,width=7.cm,height=5.5cm} & \quad %
\epsfig{figure=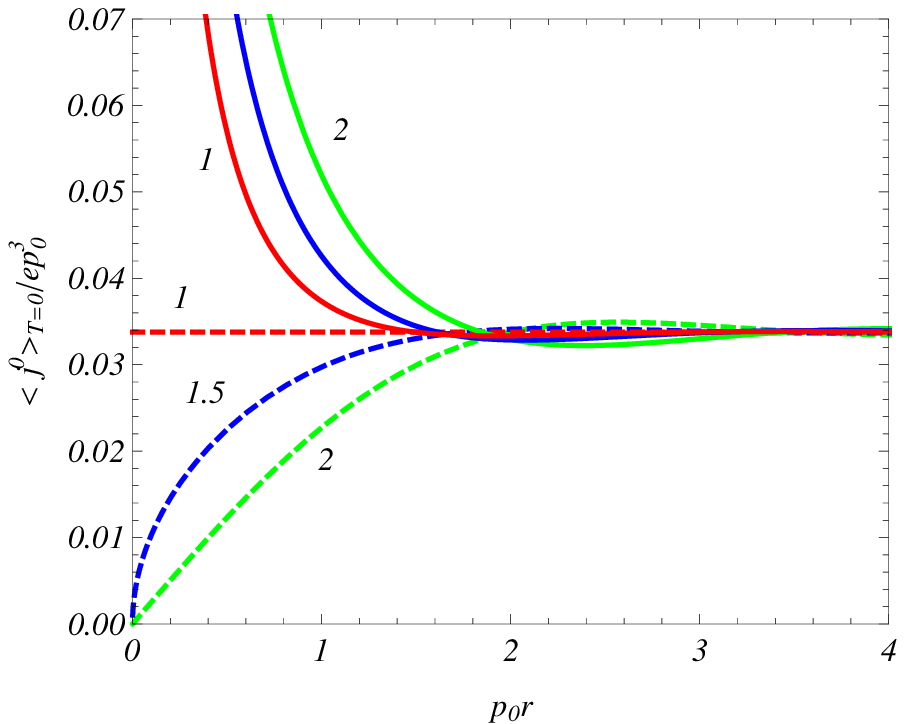,width=7.cm,height=5.5cm}%
\end{tabular}%
\end{center}
\caption{The charge density versus the radial coordinate. The full and
dashed lines correspond to the values $\protect\alpha _{0}=1/2$ and $\protect%
\alpha _{0}=0$, respectively. The numbers near the curves present the values
of the parameter $q$. The graphs on the left panel are plotted for $T/m=1$
and $\protect\mu /m=0.5$. The graphs on the right panel correspond to the
charge density at zero temperature.}
\label{fig2}
\end{figure}

In figure \ref{fig2}, the charge density is displayed as a function of the
radial coordinate. The full and dashed lines correspond to the values $%
\alpha _{0}=1/2$ and $\alpha _{0}=0$, respectively. The numbers near the
curves present the values of the parameter $q$. The graphs on the left panel
are plotted for $T/m=1$ and $\mu /m=0.5$. The graphs on the right panel
correspond to the charge density at zero temperature.

Let us consider the total charge, per unit length of the string, at zero
temperature, induced by the cosmic string and magnetic flux:%
\begin{equation}
\Delta Q_{T=0}=\phi _{0}\int_{0}^{\infty }dr\,r\left[ \left\langle
j^{0}\right\rangle _{T=0}-\left\langle j^{0}\right\rangle _{\mathrm{M},T=0}%
\right] .  \label{DeltQT0}
\end{equation}%
The integration over the radial coordinate is reduced to the integral of the
form $\int_{0}^{\infty }dx\,g(x)/x$. Note that the separate integrals with $%
\cos x$ and $\sin x$ from (\ref{gx}) diverge at the lower limit. In order to
overcome this difficulty, we write the integral as $\lim_{b\rightarrow
0}\int_{0}^{\infty }dx\,xg(x)/(x^{2}+b^{2})$. In this form the separate
integrals with $\cos x$ and $\sin x$ are expressed in terms of the error
function and taking the limit $b\rightarrow 0$ we can see that $%
\int_{0}^{\infty }dx\,g(x)/x=-1$. With this result, for the induced charge,
one gets%
\begin{equation}
\Delta Q_{T=0}=\frac{ep_{0}}{\pi q}h_{2}(q,\alpha _{0}),  \label{DeltQT0b}
\end{equation}%
where the function $h_{n}(q,\alpha _{0})$ is defined by (\ref{hnq}).

The expression (\ref{DeltQT0b}) can also be obtained by substituting the
initial expression (\ref{j0bT0}) into (\ref{DeltQT0}). Though the integrals
with separate terms in the square brackets of (\ref{DeltQT0}) diverge, the
induced total charge is finite. For the evaluation of the integral it is
convenient to take for $\left\langle j^{0}\right\rangle _{\mathrm{M},T=0}$
the expression which is obtained from (\ref{j0bT0}) taking $q=1$ and $\alpha
=0$. In order to have a possibility to integrate the terms in (\ref{DeltQT0}%
) separately, we write the radial integral as $\int_{0}^{\infty
}dr\,=\lim_{\sigma \rightarrow 0}\int_{0}^{\infty }dr\,e^{-\sigma r^{2}}$.
Then, the resulting integrals are of the form (\ref{Int-reg}) and the charge
is expressed in terms of the function (\ref{Fq}). By using the formula (\ref%
{Fq1}) for the latter, we can see that the subtraction of the Minkowskian
part cancels the contribution of the first term in the square brackets of (%
\ref{Fq1}). The remaining integral over $x$ is elementary and one gets the
result (\ref{DeltQT0b}).

In the discussion above we have considered the charge density as a function
of the chemical potential and temperature. From the physical point of view
it is of interest also to consider a system with fixed value of the charge.
With a fixed charge, the expressions presented above give the chemical
potential as a function of the temperature. At high temperatures one has $%
\left\langle j^{0}\right\rangle \approx \left\langle j^{0}\right\rangle _{%
\mathrm{M}}\approx e\mu T^{2}/3$ and, hence, for a fixed charge the chemical
potential decays as $1/T^{2}$. With decreasing temperature the function $%
|\mu (T)|$ increases. Its limiting value at zero temperature is obtained
from the expressions for the total charge at zero temperature given above.

\section{Azimuthal current}

\label{sec:Current}

Now we turn to the investigation of the current density. The only nonzero
component corresponds to the azimuthal current ($\nu =2$ in (\ref{C1})). By
taking into account the expression for the mode functions, from (\ref{jpm})
for the physical components of the current densities of the particles and
antiparticles, $\left\langle j_{\phi }\right\rangle _{\pm }=r\left\langle
j^{2}\right\rangle _{\pm }$, we get%
\begin{equation}
\left\langle j_{\phi }\right\rangle _{\pm }=\frac{eq}{4\pi ^{2}}\sum_{\sigma
}\epsilon _{j}\frac{\lambda ^{2}}{E}\frac{J_{\beta _{j}}(\lambda r)J_{\beta
_{j}+\epsilon _{j}}(\lambda r)}{e^{\beta (E\mp \mu )}+1},  \label{j2pm}
\end{equation}%
where the upper and lower signs correspond to the particles and
antiparticles respectively and the collective summation is defined by (\ref%
{Sumsig}). Note that for $\mu =0$ the contributions to the total current
density from the particles and antiparticles coincide. The current density
is an odd periodic function of the magnetic flux with the period equal to
the quantum flux.

For the case $|\mu |<m$, again we use the expansion (\ref{Expans}). This
results in the expression
\begin{equation}
\left\langle j_{\phi }\right\rangle _{\pm }=-\frac{eq}{\pi ^{2}}\
\sum_{n=1}^{\infty }(-1)^{n}\sum_{j}\ \epsilon _{j}\int_{0}^{\infty
}d\lambda \,\lambda ^{2}J_{\beta _{j}}(\lambda r)J_{\beta _{j}+\epsilon
_{j}}(\lambda r)\int_{0}^{\infty }dk\ \frac{e^{-n\beta (E\mp \mu )}}{E}.
\label{j2pm2}
\end{equation}%
By using the integral representation%
\begin{equation}
\frac{e^{-n\beta E}}{E}=\frac{2}{\sqrt{\pi }}\int_{0}^{\infty
}ds\,e^{-(\lambda ^{2}+k^{2}+m^{2})s^{2}-n^{2}\beta ^{2}/(4s^{2})},
\label{IntRep3}
\end{equation}%
the integration over $k$ is elementary. The integration over $\lambda $ is
done with the help of the formula \cite{Grad}
\begin{equation}
\int_{0}^{\infty }d\lambda \lambda ^{2}e^{-\lambda ^{2}s^{2}}J_{\beta
_{j}}(\lambda r)J_{\beta _{j}+\epsilon _{j}}(\lambda r)=r\epsilon _{j}\frac{%
e^{-x}}{4s^{4}}\ \left[ I_{\beta _{j}}(x)-I_{\beta _{j}+\epsilon _{j}}(x)%
\right] \ ,  \label{Int1}
\end{equation}%
where $x=r^{2}/(2s^{2})$. As a result, the current density is expressed as%
\begin{equation}
\left\langle j_{\phi }\right\rangle _{\pm }=-\frac{eq}{2\pi ^{2}r^{3}}%
\sum_{n=1}^{\infty }(-1)^{n}e^{\mp n\beta \mu }\int_{0}^{\infty
}dx\,xe^{-m^{2}r^{2}/(2x)-(1+n^{2}\beta ^{2}/(2r^{2}))x}G(q,\alpha _{0},x),
\label{j2pm3}
\end{equation}%
with the notation%
\begin{equation}
G(q,\alpha _{0},x)=\sum_{j}\ \left[ I_{\beta _{j}}(x)-I_{\beta _{j}+\epsilon
_{j}}(x)\right] .  \label{Gq}
\end{equation}

An equivalent expression for the current density is obtained from (\ref%
{j2pm3}) by making use of the integral representation%
\begin{eqnarray}
G(q,\alpha _{0},x) &=&\frac{4}{q}\sideset{}{'}{\sum}%
_{k=1}^{[q/2]}(-1)^{k}s_{k}\sin \left( 2\pi k\alpha _{0}\right) e^{x\cos
(2\pi k/q)}  \notag \\
&&+\frac{4}{\pi }\int_{0}^{\infty }dy\ \frac{g(q,\alpha _{0},2y)\cosh y}{%
\cosh (2qy)-\cos (q\pi )}e^{-x\cosh {(2y)}},  \label{Gq1}
\end{eqnarray}%
with the notation%
\begin{equation}
g(q,\alpha _{0},x)=\sum_{\chi =\pm 1}\chi \cos \left[ \left( 1/2+\chi \alpha
_{0}\right) q\pi \right] \cosh \left[ \left( 1/2-\chi \alpha _{0}\right) qx%
\right] \ .  \label{g}
\end{equation}
The prime on the summation sign in (\ref{Gq1}) means that, in the case where
$q$ is an even number, the term with $k=q/2$ should be taken with the
coefficient 1/2. Formula (\ref{Gq1}) is obtained by using the representation
for the series $\sum_{j}I_{\beta _{j}}(x)$ given in \cite{Beze10b}.
Substituting (\ref{Gq1}) into (\ref{j2pm3}), after integrating over $x$, we
obtain
\begin{align}
\left\langle j_{\phi }\right\rangle _{\pm }& =-\frac{4em^{4}r}{\pi ^{2}}%
\sum_{n=1}^{\infty }(-1)^{n}e^{\mp n\beta \mu }  \notag \\
& \times \left[ \ \sideset{}{'}{\sum}_{k=1}^{[q/2]}(-1)^{k}s_{k}\sin \left(
2\pi k\alpha _{0}\right) f_{2}\left( m\beta s_{n}(r/\beta ,k/q)\right)
\right.  \notag \\
& +\left. \frac{q}{\pi }\int_{0}^{\infty }dy\ \frac{\cosh \left( y\right)
g(q,\alpha _{0},2y)}{\cosh (2qy)-\cos (q\pi )}f_{2}(m\beta c_{n}(r/\beta ,y))%
\right] ,  \label{j2pm4}
\end{align}%
where the functions in the arguments of $f_{2}(x)$ are defined by (\ref{sncn}%
). In the absence of the conical defect, $q=1$, the above expression reduces
to
\begin{align}
\left\langle j_{\phi }\right\rangle _{\pm }& =\frac{4em^{4}r}{\pi ^{3}}\sin
(\alpha _{0}\pi )\sum_{n=1}^{\infty }(-1)^{n}e^{\mp n\beta \mu }  \notag \\
& \times \int_{0}^{\infty }dy\,\cosh (2\alpha _{0}y)f_{2}\left( m\beta
c_{n}(r/\beta ,y)\right) .  \label{j2pmq1}
\end{align}

Now, by taking into account the expression for the vacuum expectation value
of the current density from \cite{Beze13}, the total current density is
expressed as
\begin{align}
\left\langle j_{\phi }\right\rangle & =-\frac{8em^{4}r}{\pi ^{2}}%
\sideset{}{'}{\sum}_{n=0}^{\infty }(-1)^{n}\cosh (n\beta \mu )  \notag \\
& \times \left[ \ \sideset{}{'}{\sum}_{k=1}^{[q/2]}(-1)^{k}s_{k}\sin \left(
2\pi k\alpha _{0}\right) f_{2}\left( m\beta s_{n}(r/\beta ,k/q)\right)
\right.  \notag \\
& +\left. \frac{q}{\pi }\int_{0}^{\infty }dy\ \frac{\cosh \left( y\right)
g(q,\alpha _{0},2y)}{\cosh (2qy)-\cos (q\pi )}f_{2}\left( m\beta
c_{n}(r/\beta ,y)\right) \right] ,  \label{j2}
\end{align}%
where the prime on the sign of the summation over $n$ means that the term $%
n=0$ should be taken with the coefficient 1/2. This term corresponds to the
vacuum expectation value of the current density, $\left\langle j_{\phi
}\right\rangle _{0}$.

For a massless field, because of the condition $\mu \leqslant m$, we should
also take $\mu =0$. By using the asymptotic expression for the MacDonald
function for small values of the argument, the summation over $n$ is reduced
to the series of the form $\sideset{}{'}{\sum}_{n=0}^{\infty }(-1)^{n}\left(
n^{2}+x^{2}\right) ^{-2}$. The sum of the latter is expressed in terms of
the hyperbolic functions and one gets%
\begin{align}
\left\langle j_{\phi }\right\rangle & =-\frac{eT}{2\pi r^{2}}\left[ \ %
\sideset{}{'}{\sum}_{k=1}^{[q/2]}\frac{(-1)^{k}}{s_{k}^{2}}\sin \left( 2\pi
k\alpha _{0}\right) h(2rTs_{k})\right.  \notag \\
& \left. +\frac{q}{\pi }\int_{0}^{\infty }dy\ \frac{g(q,\alpha _{0},2y)}{%
\cosh (2qy)-\cos (q\pi )}\frac{h(2rT\cosh y)}{\cosh ^{2}y}\right] \ ,
\label{j2pmm0}
\end{align}%
where we have introduced the function%
\begin{equation}
h(x)=\frac{1+\pi x\coth (\pi x)}{\sinh (\pi x)}.  \label{hx}
\end{equation}%
For $|x|\ll 1$ one has $h(x)\approx 2/(\pi x)-7(\pi x)^{3}/180$. The
contribution corresponding to the first term in the right-hand side of the
latter expression gives the vacuum expectation value of the current density:%
\begin{equation}
\left\langle j_{\phi }\right\rangle _{0}=-\frac{eg_{4}(q,\alpha _{0})}{2\pi
^{2}r^{3}},  \label{j2vev}
\end{equation}%
where we have defined%
\begin{equation}
g_{n}(q,\alpha _{0})=\sideset{}{'}{\sum}_{k=1}^{[q/2]}(-1)^{k}\frac{\sin
\left( 2\pi k\alpha _{0}\right) }{s_{k}^{n-1}}+\frac{q}{\pi }%
\int_{0}^{\infty }dy\ \frac{g(q,\alpha _{0},2y)\cosh ^{1-n}y}{\cosh
(2qy)-\cos (q\pi )}.  \label{gn}
\end{equation}%
The thermal part in the current density for a massless field is given by the
right-hand side of (\ref{j2pmm0}) with the replacement $h(x)\rightarrow
h(x)-2/(\pi x)$. This part vanishes on the string. Hence, near the string, $%
rT\ll 1$, the total current is dominated by the part $\left\langle j_{\phi
}\right\rangle _{0}$.

Now we turn to the investigation of the current density in asymptotic
regions of the parameters. First we consider the case of a massless field.
For $rT\ll 1$, from (\ref{j2pmm0}), to the leading order one has $%
\left\langle j_{\phi }\right\rangle \approx \left\langle j_{\phi
}\right\rangle _{0}$, with $\left\langle j_{\phi }\right\rangle _{0}$ given
by (\ref{j2vev}). This limit corresponds to points close to the string, for
a fixed value of the temperature, or to lower temperatures, for fixed $r$.
The relative contribution of the thermal effects is suppressed by the factor
$(rT)^{4}$. In the opposite limit, $rT\gg 1$, two cases should be considered
separately. For $q<2$, in (\ref{j2pmm0}) the sum over $k$ is absent and the
dominant contribution to the integral comes from values of $y$ near the
lower limit. Assuming that $\sin (q\pi /2)\gg q/\sqrt{\pi rT}$ ($q$ is not
too close to 2), to the leading order one gets%
\begin{equation}
\left\langle j_{\phi }\right\rangle \approx \frac{eq(rT)^{3/2}\sin (\alpha
_{0}q\pi )}{\pi r^{3}\sin (q\pi /2){e}^{2\pi rT}}.  \label{j2m0HighT}
\end{equation}%
In the case $q>2$, the contribution of the term with $k=1$ dominates and we
find%
\begin{equation}
\left\langle j_{\phi }\right\rangle \approx \frac{2e\sin \left( 2\pi \alpha
_{0}\right) }{r^{3}\sin {(\pi /q)}}\frac{(rT)^{2}}{{e}^{2\pi rT\sin {(\pi /q)%
}}}.  \label{j2m0HighTb}
\end{equation}%
In both cases the current density is exponentially suppressed. The
expressions (\ref{j2m0HighT}) and (\ref{j2m0HighTb}) correspond to high
temperatures for fixed $r$ or to large distances from the string for fixed $%
T $.

For a massive field, in the limit $r\rightarrow 0$ and for $2|\alpha
_{0}|<1-1/q$, in the expression of the square brackets in (\ref{j2}),
corresponding to the finite temperature terms, we can directly put $r=0$:%
\begin{equation}
\left\langle j_{\phi }\right\rangle \approx \left\langle j_{\phi
}\right\rangle _{0}-\frac{8em^{4}r}{\pi ^{2}}g_{0}(q,\alpha
_{0})\sum_{n=1}^{\infty }(-1)^{n}\cosh (n\beta \mu )f_{2}\left( nm\beta
\right) .  \label{j2r0}
\end{equation}%
In the limit $r\rightarrow 0$, for the vacuum expectation value, to the
leading order, one has%
\begin{equation}
\left\langle j_{\phi }\right\rangle _{0}\approx -\frac{eg_{4}(q,\alpha _{0})%
}{2\pi ^{2}r^{3}}.  \label{j20r0}
\end{equation}%
The latter diverges on the string and, hence, near the string the total
current is dominated by the zero temperature part.

In the case $2|\alpha _{0}|>1-1/q$ and in the limit $r\rightarrow 0$, the
integral over $y$ in (\ref{j2}), corresponding to the the thermal part ($%
n\neq 0$), diverges and we cannot directly put $r=0$. We note that in this
case the dominant contribution to the integral over $y$ comes from large
values of $y$. Replacing the integrand by its asymptotic form for large $y$,
the integral is evaluated in terms of the MacDonald function and one gets%
\begin{eqnarray}
\left\langle j_{\phi }\right\rangle &\approx &\left\langle j_{\phi
}\right\rangle _{0}+\frac{2^{3/2}}{\pi ^{2}}\mathrm{sgn}(\alpha _{0})\frac{%
eqm^{3}(mr/\sqrt{2})^{\left( 1-2|\alpha _{0}|\right) q}}{\Gamma (1/2+\left(
1/2-|\alpha _{0}|\right) q)}  \notag \\
&&\times \sum_{n=1}^{\infty }(-1)^{n}\cosh (n\beta \mu )f_{3/2+\left(
1/2-|\alpha _{0}|\right) q}(nm\beta ).  \label{j2r0b}
\end{eqnarray}%
For the vacuum expectation value we have the asymptotic expression (\ref%
{j20r0}) and, again, it dominates for points near the string.

Now let us discuss the asymptotic expressions for the current density at low
and high temperatures. At low temperatures, $T\ll m,r^{-1}$, and for $%
2|\alpha _{0}|<1-1/q$, in (\ref{j2}), for the terms corresponding to the
thermal corrections we can put $r/\beta =0$ in the arguments of the
functions $f_{2}(x)$. By using the asymptotic expression of the MacDonald
function for large arguments we find%
\begin{equation}
\left\langle j_{\phi }\right\rangle \approx \left\langle j_{\phi
}\right\rangle _{0}+\frac{4em^{4}re^{-\left( m-|\mu |\right) \beta }}{\sqrt{2%
}\pi ^{3/2}\left( m\beta \right) ^{5/2}}g_{0}(q,\alpha _{0}).  \label{j2T0}
\end{equation}%
In the case $2|\alpha _{0}|\geqslant 1-1/q$ and for low temperatures we
cannot directly put $r/\beta =0$ in the integrand of (\ref{j2}) for the
thermal part because the resulting integral is divergent. At low
temperatures the integral is dominated by the contribution of large $y$. By
using the asymptotic expression for the integrand, we get
\begin{equation}
\left\langle j_{\phi }\right\rangle \approx \left\langle j_{\phi
}\right\rangle _{0}-\frac{eqm(mr/\sqrt{2})^{\left( 1-2|\alpha _{0}|\right) q}%
}{\pi ^{3/2}\Gamma (1/2+\left( 1/2-|\alpha _{0}|\right) q)}\frac{\mathrm{sgn}%
(\alpha _{0})e^{-(m-|\mu |)\beta }}{\beta ^{2}(m\beta )^{\left( 1/2-|\alpha
_{0}|\right) q}}.  \label{j2T0b}
\end{equation}%
Hence, under the condition $|\mu |<m$, at low temperatures the thermal
corrections are exponentially suppressed.

In order to investigate the high temperature asymptotic behavior, it is
convenient to present the current density (\ref{j2pm4}) in an alternative
form by using the relation (\ref{SumHighT}). The corresponding expression
reads%
\begin{eqnarray}
\left\langle j_{\phi }\right\rangle  &=&-\frac{16eTr}{(2\pi )^{3/2}}%
\sum_{n=-\infty }^{+\infty }b_{n}^{3}(T)\left[ \ \sideset{}{'}{\sum}%
_{k=1}^{[q/2]}(-1)^{k}s_{k}\sin \left( 2\pi k\alpha _{0}\right)
f_{3/2}(2rs_{k}b_{n}(T))\right.   \notag \\
&&\left. +\frac{q}{\pi }\int_{0}^{\infty }dy\ \frac{\cosh \left( y\right)
g(q,\alpha _{0},2y)}{\cosh (2qy)-\cos (q\pi )}f_{3/2}(2r\cosh yb_{n}(T))%
\right] ,  \label{j2pm5}
\end{eqnarray}%
where $b_{n}(T)$ is defined by (\ref{bn}). Note that $f_{3/2}(x)=\sqrt{\pi /2%
}x^{-3}(1+x)e^{-x}$. In (\ref{j2pm5}) we can also write $\sum_{n=-\infty
}^{+\infty }\cdots =2\sum_{n=0}^{\infty }\mathrm{Re}\left[ \cdots \right] $.
For a massless fermionic field, because of the condition $|\mu |\leqslant m$%
, we should also take $\mu =0$ and one gets%
\begin{align}
\left\langle j_{\phi }\right\rangle & =-\frac{eT}{\pi r^{2}}%
\sum_{n=0}^{\infty }\left[ \ \sideset{}{'}{\sum}_{k=1}^{[q/2]}(-1)^{k}\sin
\left( 2\pi k\alpha _{0}\right) \frac{1+2\pi (2n+1)s_{k}Tr}{s_{k}^{2}e^{2\pi
(2n+1)s_{k}Tr}}\right.   \notag \\
& +\left. \frac{q}{\pi }\int_{0}^{\infty }dy\ \frac{g(q,\alpha _{0},2y)}{%
\cosh (2qy)-\cos (q\pi )}\frac{1+2\pi (2n+1)Tr\cosh y}{e^{2\pi (2n+1)Tr\cosh
y}\cosh ^{2}y}\right] .  \label{j2pm5m0}
\end{align}

At high temperatures one has $r|b_{n}(T)|\gg 1$, and the dominant
contribution in (\ref{j2pm5}) comes from the terms $n=0$ and $n=-1$. For $%
q>2 $ the leading contribution comes from the $k=1$ term in the right-hand
side and we get%
\begin{equation}
\left\langle j_{\phi }\right\rangle \approx \ \frac{2eT^{2}\sin \left( 2\pi
\alpha _{0}\right) }{r\sin {(\pi /q)}}\cos (2\mu r\sin {(\pi /q)})e^{-2\pi
rT\sin {(\pi /q)}}\ .  \label{j2HighT1}
\end{equation}%
For $q<2$, the sum over $k$ in (\ref{j2pm5}) is absent. At high
temperatures, the dominant contribution to the integral over $y$ comes from
the region near the lower limit of the integration. Assuming that $\sin (\pi
q/2)\gg 1/\sqrt{rT}$ ($q$ is not too close to 2), to the leading order one
finds%
\begin{equation}
\left\langle j_{\phi }\right\rangle \approx \frac{eq\sin (\alpha _{0}q\pi
)T^{3/2}}{\pi \sin (\pi q/2)r^{3/2}}\cos (2\mu r)e^{-2\pi rT}.
\label{j2HighT2}
\end{equation}%
Hence, at high temperatures, $T\gg m,r^{-1}$, the current density is
exponentially small.

At large distances from the string, in (\ref{j2pm5}) the contribution from
the terms with $k=1$ and $n=-1,0$ dominate and for $q>2$ one gets%
\begin{equation}
\left\langle j_{\phi }\right\rangle \approx \frac{2eT}{\pi r}\frac{\sin
\left( 2\pi k\alpha _{0}\right) }{\sin (\pi /q)}\mathrm{Re\,}[b_{0}(T)\
e^{-2rb_{0}(T)\sin (\pi /q)}].  \label{j2Larger}
\end{equation}%
For $q<2$ the sum over $k$ in (\ref{j2pm5}) is absent and at large distances
the dominant contribution comes from the region near the lower limit of the
integral over $y$. Assuming that $rT\gg \sin ^{-2}(\pi q/2)$, to the leading
order we find
\begin{equation}
\left\langle j_{\phi }\right\rangle \approx \frac{eqT}{\pi ^{3/2}r^{2}}\frac{%
\sin (\pi q\alpha _{0})}{\sin (\pi q/2)}\mathrm{Re\,}%
[(rb_{0}(T))^{1/2}e^{-2rb_{0}(T)}]\ .  \label{j2Largerb}
\end{equation}%
In both cases one has an exponential decay.

\begin{figure}[tbph]
\begin{center}
\begin{tabular}{cc}
\epsfig{figure=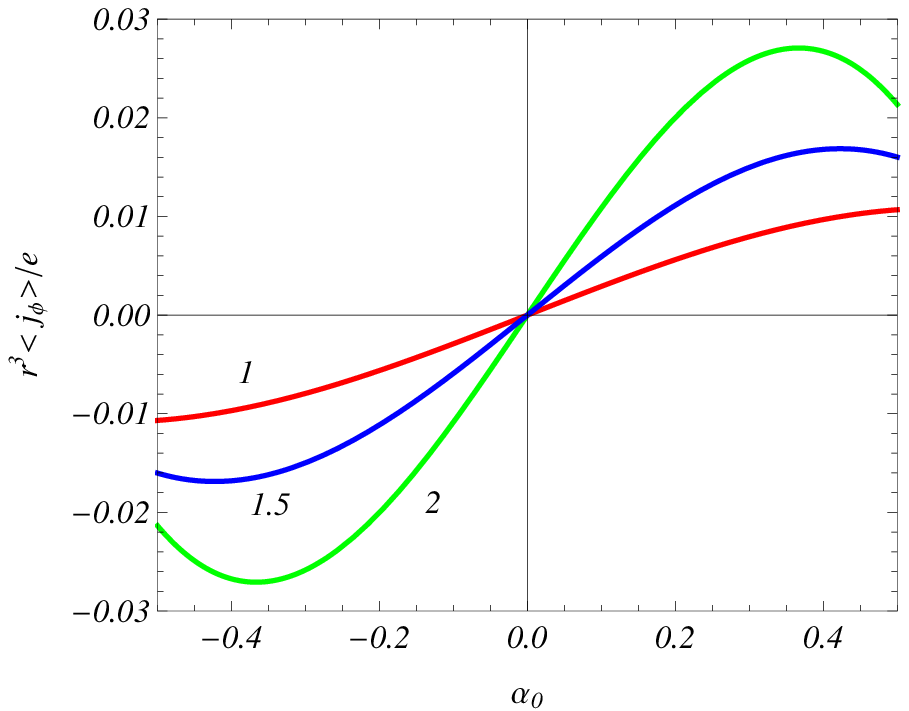,width=7.cm,height=5.5cm} & \quad %
\epsfig{figure=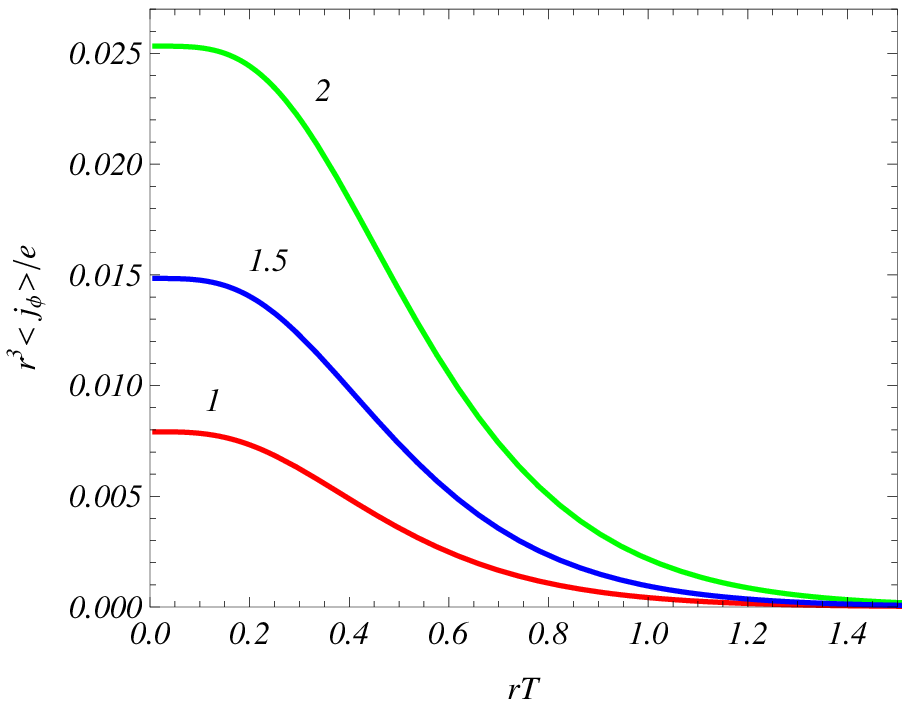,width=7.cm,height=5.5cm}%
\end{tabular}%
\end{center}
\caption{The azimuthal current density for a massless field with zero
chemical potential as a function of $\protect\alpha _{0}$ (left panel) and
as a function of the temperature (right panel). The numbers near the curves
correspond to the values of $q$. For the graphs on the left panel $rT=0.25$
and for the right panel $\protect\alpha _{0}=0.25$.}
\label{fig3}
\end{figure}

In figure \ref{fig3}, for a massless field with $\mu =0$, we have plotted
the azimuthal current density as a function of the parameter $\alpha _{0}$
(left panel) and as a function of the temperature (right panel). The numbers
near the curves correspond to the values of $q$. The graphs on the left
panel are plotted for $rT=0.25$ and on the right panel for $\alpha _{0}=0.25$%
.

Now let us consider the azimuthal current density in the case $|\mu |>m$.
The current density is an even function of the chemical potential and it
suffices to consider positive values of $\mu $. In this case the
contribution of the antiparticles to the current density is evaluated in a
way similar to that we have described for $\mu <m$. This contribution is
given by (\ref{j2pm4}) with the lower sign. For the total current one gets%
\begin{equation}
\left\langle j_{\phi }\right\rangle =\left\langle j_{\phi }\right\rangle
_{0}+\left\langle j_{\phi }\right\rangle _{-}+\frac{eq}{\pi ^{2}}%
\int_{0}^{\infty }dk\int_{0}^{\infty }d\lambda \ \sum_{j}\epsilon _{j}\frac{%
\lambda ^{2}}{E}\frac{J_{\beta _{j}}(\lambda r)J_{\beta _{j}+\epsilon
_{j}}(\lambda r)}{e^{\beta (E-\mu )}+1},  \label{j2b}
\end{equation}%
where the last term is the part coming from the particles. At high
temperatures, $T\gg \mu $, the dominant contribution comes from the states
with energies $E\gg \mu $ and the asymptotic behavior of the current density
is similar to that we have described above for the case $\mu <m$. At zero
temperature, taking the limit $T\rightarrow 0$ in (\ref{j2b}), we obtain%
\begin{equation}
\left\langle j_{\phi }\right\rangle _{T=0}=\left\langle j_{\phi
}\right\rangle _{0}+\frac{eq}{\pi ^{2}}\int_{0}^{p_{0}}du\,\frac{u^{3}}{%
\sqrt{u^{2}+m^{2}}}\int_{0}^{1}dx\frac{x^{2}}{\sqrt{1-x^{2}}}%
\sum_{j}\epsilon _{j}J_{\beta _{j}}(urx)J_{\beta _{j}+\epsilon _{j}}(urx).
\label{j2bT0}
\end{equation}%
The second term in the right-hand side is the contribution from the
particles filling the states with the energies $E\leqslant \mu $.

For the further transformation of the zero temperature current density, in (%
\ref{j2bT0}) we use the relation%
\begin{equation}
\epsilon _{j}J_{\beta _{j}}(z)J_{\beta _{j}+\epsilon _{j}}(z)=\frac{1}{z}%
\left( \epsilon _{j}\beta _{j}-\frac{1}{2}z\partial _{z}\right) J_{\beta
_{j}}^{2}(z),  \label{RelJ}
\end{equation}%
and the integral representation (\ref{JRep}) for the square of the Bessel
function. Next, by taking into account that
\begin{equation}
\left( \epsilon _{j}\beta _{j}-z\partial _{z}/2\right) e^{-z}I_{\beta
_{j}}(z)=ze^{-z}\left[ I_{\beta _{j}}(z)-I_{\beta _{j}+\epsilon _{j}}(z)%
\right] ,  \label{RelI}
\end{equation}%
the series over $j$ is expressed in terms of the function (\ref{Gq}). By
using the representation for the latter given by (\ref{Gq1}), the integral
over $t$ coming from (\ref{JRep}) is expressed in terms of the Bessel
function $J_{1}(2urxb)$, where $b=s_{k}$ and $b=\cosh y$ for the first and
second terms in the right-hand side of (\ref{Gq1}), respectively. After the
integration over $x$, we find%
\begin{eqnarray}
\left\langle j_{\phi }\right\rangle _{T=0} &=&\left\langle j_{\phi
}\right\rangle _{0}-\frac{ep_{0}^{3}}{\pi ^{2}}\left[ \sideset{}{'}{\sum}%
_{k=1}^{[q/2]}(-1)^{k}\sin \left( 2\pi k\alpha _{0}\right)
f(p_{0}rs_{k},m/p_{0})\right.   \notag \\
&&\left. +\frac{q}{\pi }\int_{0}^{\infty }dy\ \frac{f(p_{0}r\cosh y,m/p_{0})%
}{\cosh (2qy)-\cos (q\pi )}g(q,\alpha _{0},2y)\right] ,  \label{j2bT01}
\end{eqnarray}%
with the notation
\begin{equation}
f(a,b)=\frac{1}{a}\int_{0}^{1}dx\,\frac{x^{2}g(2ax)}{\sqrt{x^{2}+b^{2}}}.
\label{fab}
\end{equation}%
Here, the function $g(x)$ is defined by (\ref{gx}). For a massless field one
has
\begin{equation}
f(a,0)=a^{-3}\sin ^{2}a\left( a\cot a-1\right) .  \label{fa0}
\end{equation}

The second term in the right-hand side of (\ref{j2bT01}) vanishes on the
string, whereas the first one diverges as $1/r^{3}$. Hence, near the string
the current density is dominated by the vacuum expectation value. In order
to find the asymptotic behavior of (\ref{j2bT01}) at large distances, $%
p_{0}r\gg 1$, we note that for $a\gg 1$ one has%
\begin{equation}
f(a,b)\approx \frac{1}{2a^{2}}\frac{\sin (2a)}{\sqrt{1+b^{2}}}.
\label{fabLarge}
\end{equation}%
Hence, at large distances, the contribution in the current density coming
form the particles exhibits a damping oscillatory behavior with the
amplitude decaying as $1/r^{2}$. The vacuum expectation value decays as $%
1/r^{3}$ for a massless field and exponentially for a massive field and,
hence, it is subdominant at large distances.

\begin{figure}[tbph]
\begin{center}
\epsfig{figure=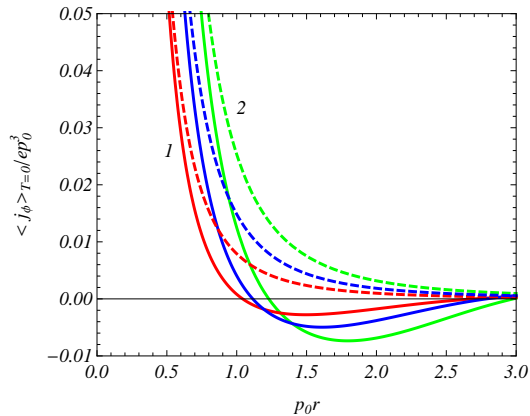,width=7.cm,height=5.5cm}
\end{center}
\caption{The zero temperature azimuthal current density for a massless field
with $\protect\mu >0 $, as a function of the radial coordinate (full
curves). The numbers near the curves correspond to the values of $q$. The
graphs are plotted for $\protect\alpha _{0}=0.25$. The dashed curves
correspond to the vacuum expectation value of the azimuthal current.}
\label{fig4}
\end{figure}

In figure \ref{fig4}, for a massless field, we have plotted the azimuthal
current density at zero temperature and $\mu >0$ (full curves), as a
function of the radial coordinate, for various values of the parameter $q$
(numbers near the curves, the curve in the middle corresponds to $q=1.5$).
For the value of $\alpha _{0}$ we have taken $\alpha _{0}=0.25$. The dashed
curves correspond to the vacuum expectation value of the current density, $%
\left\langle j_{\phi }\right\rangle _{0}$, for the same values of $q$.

\section{Conclusion}

\label{sec:Conc}

In this paper, we have investigated the combined effects of the planar angle
deficit and the magnetic flux on the charge and current densities for a
massive fermionic field with nonzero chemical potential at thermal
equilibrium. For the evaluation of the corresponding expectation values the
direct summation over a complete set of fermionic modes is used. These
densities are decomposed into the vacuum expectation values and finite
temperature contributions, coming from the particles and antiparticles.

For the charge density the vacuum expectation value vanishes and the
expectation value for the particles and antiparticles in the case $|\mu
|\leqslant m$ are given by (\ref{j0pm3}), where $\left\langle
j^{0}\right\rangle _{\mathrm{M}\pm }$ is the corresponding quantity in the
absence of the planar angle deficit and the magnetic flux. The charge
density is an even periodic function of the magnetic flux with period equal
to the quantum flux. For the zero chemical potential the contributions from
the particles and antiparticles cancel each other and the total charge
density, given by (\ref{j0}), vanishes. An alternative expression for the
charge density, convenient for the investigation of high temperature regime,
is provided by (\ref{j0HighT}). The charge density is finite on the string
for the values of the magnetic flux in the range $2|\alpha _{0}|\leqslant
1-1/q$ and diverges as $r^{\left( 1-2|\alpha _{0}|\right) q-1}$ for $%
2|\alpha _{0}|>1-1/q$. This divergence is integrable and the total charge
induced by the cosmic string and magnetic flux is finite (see (\ref{DelQ})).
At large distances from the string and for $|\mu |\leqslant m$, the
asymptotic expressions for the charge density in the cases $q>2$ and $q<2$
are given by (\ref{j0Larger}) and (\ref{j0Largerb}), respectively. In this
limit, the contributions in the charge density induced by the string and
magnetic flux are exponentially small. At low temperatures the charge
density is suppressed by the factor $e^{-\beta (m-|\mu |)}$. At high
temperatures, the contribution to the charge density coming from the string
and magnetic flux is suppressed by the factor $e^{-2\pi rT\sin (\pi /q)}$
for $q>2$ and by the factor $e^{-2\pi rT}$ for $q<2$. In this limit the
charge density is dominated by the Minkowskian part with the high
temperature asymptotic given by (\ref{j0MhighT}).

We have also investigated the charge density for the case $|\mu |>m$. By
taking into account that the charge density is an odd function of the
chemical potential, for definiteness, it is assumed that $\mu >0$. At high
temperatures, the dominant contribution to the expectation values comes from
the states with the energies $E\gg |\mu |$ and the corresponding behavior is
similar to that for the case $\mu <m$. The behavior of the charge density is
essentially changed at low temperatures. Now, in the limit $T\rightarrow 0$
the charge density does not vanish and the corresponding limiting value is
given by (\ref{j0bT0}). The appearance of the nonzero charge density at zero
temperature is related to the presence of particles filling the states with
the energies $m\leqslant E\leqslant \mu $. An alternative expression for the
charge density at zero temperature is provided by (\ref{j0bT01}) where the
Minkowskian part is given by (\ref{j0MbT0}). The zero temperature charge
density is finite on the string for $2|\alpha _{0}|\leqslant 1-1/q$ and
diverges for $2|\alpha _{0}|>1-1/q$. The divergence is integrable and the
total charge induced by the cosmic string and magnetic flux is finite and is
given by the expression (\ref{DeltQT0b}). At large distances from the
string, the zero temperature charge density induced by the string and
magnetic flux exhibits an oscillatory behavior with the amplitude decaying
as $1/r^{2}$. This behavior is in contrast with the case $|\mu |<m$ where we
had an exponential suppression.

The only nonzero component of the expectation value for the current density
corresponds to the current along the azimuthal direction. This current
vanishes in the absence of the magnetic flux and is an odd periodic function
of the latter with the period equal to the quantum flux. The azimuthal
current density is an even function of the chemical potential. For the zero
chemical potential, the contributions to the total current density from the
particles and antiparticles coincide. Similar to the case of the charge
density, in the case $|\mu |\leqslant m$, for the azimuthal current we have
provided two equivalent representations, (\ref{j2}) and (\ref{j2pm5}). The
first one is convenient for the investigation of the asymptotic behavior
near the string and at low temperatures, whereas the second one is well
adapted for the consideration of the asymptotics at large distances and high
temperatures. For a massless fermionic field the corresponding expressions
are simplified to (\ref{j2pmm0}) and (\ref{j2pm5m0}). The contribution in
the current density coming from particles and antiparticles vanish on the
string whereas the vacuum expectation value diverges as $r^{-3}$ and near
the string the latter dominates in the total current. At large distances
from the string, $r\gg T^{-1},m^{-1}$, and for $|\mu |<m$, the leading terms
in the asymptotic expansions over the distance from the string are given by (%
\ref{j2Larger}) and (\ref{j2Largerb}) and the current density decays
exponentially. At low temperatures, under the condition $|\mu |<m$, the
finite temperature part in the current density is suppressed by the factor $%
e^{-(m-|\mu |)/T}$ and the vacuum expectation value dominates. At high
temperatures, the leading terms in the expectation value of the current
density are given by the expressions (\ref{j2HighT1}) and (\ref{j2HighT2})
for the cases $q>2$ and $q<2$, respectively. In both cases the current
density is exponentially small. At high temperatures the contribution of the
states with the energies $E\gg \mu $ dominates and this behavior is valid
for the case $|\mu |>m$ as well. This is not the case at low temperatures.
For $|\mu |>m$ and in the limit $T\rightarrow 0$, the current density has
two contributions. The first one corresponds to the vacuum expectation value
and the second one comes from the particles in the case $\mu >0$ and from
the antiparticles for $\mu <0$, filling the states with the energies $%
m\leqslant E\leqslant |\mu |$. The current density is an even function of
the chemical potential and we have considered the case $\mu >m$. The two
alternative representations for the azimuthal current at zero temperature
are given by (\ref{j2bT0}) and (\ref{j2bT01}). The part in the zero
temperature current density coming from the particles vanishes on the string
and for points near the string the vacuum expectation value dominates. At
large distances from the string, $p_{0}r\gg 1$, the contribution in the zero
temperature current density coming form the particles exhibits a damping
oscillatory behavior with the amplitude decaying as $1/r^{2}$ and it
dominates over the vacuum current.

\section*{Acknowledgments}

The authors thank Conselho Nacional de Desenvolvimento Cient\'{\i}fico e
Tecnol\'{o}gico (CNPq) for the financial support. A. A. S. was supported by
the State Committee of Science of the Ministry of Education and Science RA,
within the frame of Grant No. SCS 13-1C040.


\begin{thebibliography}{99}
\bibitem{Nielsen} N.B. Nielsen and P. Olesen, Nucl. Phys. B \textbf{61}, 45
(1973).

\bibitem{Garfinkle} D. Garfinkle, Phys. Rev. D \textbf{32}, 1323 (1985).

\bibitem{Linet1} B. Linet, Phys. Lett. B \textbf{124}, 240 (1987).

\bibitem{Vile94} A. Vilenkin and E.P.S. Shellard, \textit{Cosmic Strings and
Other Topological Defects} (Cambridge University Press, Cambridge, 1994).

\bibitem{Sara02} S. Sarangi and S.H.H. Tye, Phys. Lett. B \textbf{536}, 185
(2002); E.J. Copeland, R.C. Myers, and J. Polchinski, JHEP \textbf{06}, 013
(2004); G. Dvali and A. Vilenkin, JCAP \textbf{0403}, 010 (2004).

\bibitem{Nels02} D.R. Nelson, \textit{Defects and Geometry in Condensed
Matter Physics} (Cambridge University Press, Cambridge, 2002); G.E. Volovik,
\textit{The Universe in a Helium Droplet} (Clarendon, Oxford, 2003).

\bibitem{Dowk87} J.S. Dowker, Phys. Rev. D \textbf{36}, 3095 (1987); J.S.
Dowker, Phys. Rev. D \textbf{36}, 3742 (1987).

\bibitem{Guim94} M.E.X. Guimar\~{a}es and B. Linet, Commun. Math. Phys.
\textbf{165}, 297 (1994).

\bibitem{Spin03} J. Spinelly and E.R. Bezerra de Mello, Class. Quantum Grav.
\textbf{20} 874, (2003); J. Spinelly and E.R. Bezerra de Mello, Int. J. Mod.
Phys. A \textbf{17}, 4375 (2002).

\bibitem{Spin04} J. Spinelly and E.R. Bezerra de Mello, Int. J. Mod. Phys. D
\textbf{13}, 607 (2004).

\bibitem{Spin08} J. Spinelly and E. R. Bezerra de Mello, JHEP \textbf{09},
005 (2008).

\bibitem{Site12} Yu.A. Sitenko and N.D. Vlasii, Class. Quantum Grav. \textbf{%
29}, 095002 (2012).

\bibitem{Alfo89} M.G. Alford and F. Wilczek, Phys. Rev. Lett. \textbf{62},
1071 (1989); K. Jones-Smith, H. Mathur, and T. Vachaspati, Phys. Rev. D
\textbf{81}, 043503 (2010); Y.-Z. Chu, H. Mathur, and T. Vachaspati, Phys.
Rev. D \textbf{82}, 063515 (2010); D.A. Steer and T. Vachaspati, Phys. Rev.
D \textbf{83}, 043528 (2011).

\bibitem{Srir01} L. Sriramkumar, Class. Quantum Grav. \textbf{18}, 1015
(2001).

\bibitem{Site09} Yu.A. Sitenko and N.D. Vlasii, Class. Quantum Grav. \textbf{%
26}, 195009 (2009).

\bibitem{Beze10} E.R. Bezerra de Mello, Class. Quantum Grav. \textbf{27},
095017 (2010).

\bibitem{Beze13} E.R. Bezerra de Mello and A. A. Saharian, Eur. Phys. J. C
\textbf{73}, 2532 (2013).

\bibitem{Brag14} E.A.F. Bragan\c{c}a, H.F. Santana Mota, and E. R. Bezerra
de Mello, arXiv:1410.1511.

\bibitem{Beze10b} E.R. Bezerra de Mello, V.B. Bezerra, A.A. Saharian, and
V.M. Bardeghyan, Phys. Rev. D \textbf{82}, 085033 (2010).

\bibitem{Bell10} S. Bellucci and A.A. Saharian, Phys. Rev. D \textbf{82},
065011 (2010); S. Bellucci and A.A. Saharian, Phys. Rev. D \textbf{87},
025005 (2013).

\bibitem{Bell13} S. Bellucci, A.A. Saharian, and H.A. Nersisyan, Phys. Rev.
D \textbf{88}, 024028 (2013).

\bibitem{Beze14} E.R. Bezerra de Mello, A.A. Saharian, and V. Vardanyan,
arXiv:1410.2860.

\bibitem{Moha14} A. Mohammadi, E.R. Bezerra de Mello, and A.A. Saharian,
arXiv:1407.8095.

\bibitem{Beze14b} E.R. Bezerra de Mello, V.B. Bezerra, A.A. Saharian, and
H.H. Harutyunyan, arXiv:1411.1258.

\bibitem{Beze13T} E.R. Bezerra de Mello and A.A. Saharian, Phys. Rev. D
\textbf{87}, 045015 (2013).

\bibitem{Bell14T} S. Bellucci, E.R. Bezerra de Mello and A.A. Saharian,
Phys. Rev. D \textbf{89}, 085002 (2014).

\bibitem{Davi88} P.C.W. Davies and V. Sahni, Class. Quantum Grav. \textbf{5}%
, 1 (1988).

\bibitem{Line92} B. Linet, Class. Quantum Grav. \textbf{9}, 2429 (1992).

\bibitem{Frol95} V.P. Frolov, A. Pinzul and A.I. Zelnikov, Phys. Rev. D
\textbf{51}, 2770 (1995).

\bibitem{Guim95} M.E.X. Guimar\~{a}es, Class. Quantum Grav. \textbf{12},
1705 (1995).

\bibitem{Line96} B. Linet, Class. Quantum Grav. \textbf{13}, 97 (1996).

\bibitem{Sous89} P. de Sousa Gerbert and R. Jackiw, Commun. Math. Phys.
\textbf{124}, 229 (1989); P. de Sousa Gerbert, Phys. Rev. D \textbf{40},
1346 (1989); Yu.A. Sitenko, Ann. Phys. \textbf{282}, 167 (2000).

\bibitem{Jack09} R. Jackiw, A.I. Milstein, S.-Y. Pi, and I.S. Terekhov,
Phys. Rev. B \textbf{80}, 033413 (2009); A.I. Milstein and I.S. Terekhov,
Phys. Rev. B \textbf{83}, 075420 (2011).

\bibitem{Grad} I.S. Gradshteyn and I.M. Ryzhik, \textit{Table of Integrals,
Series and Products} (Academic Press, New York, 1980).

\bibitem{Bell09} S. Bellucci and A.A. Saharian, Phys. Rev. D \textbf{79},
085019 (2009).

\bibitem{Wats66} G.N. Watson, \textit{A Treatise on the Theory of Bessel
Functions} (Cambridge University Press, Cambridge, 1966).
\end{thebibliography}
\end{document}